\documentclass[trackchanges]{aastex63}
\usepackage[figuresright]{rotating}
\usepackage{multirow}

\begin{document}

\title{Photometric and spectroscopic studies of the long period low mass ratio deep contact binary KN Per}

\correspondingauthor{Kai Li}
\email{kaili@sdu.edu.cn}

\author{Xin-Yi Gao}
\affiliation{SDU-ADU Joint Science College, Shandong University, Weihai, Shandong, 264209, China.}

\author{Kai Li}
\affiliation{Shandong Key Laboratory of Optical Astronomy and Solar-Terrestrial Environment,\\
 School of Space Science and Physics, \\
 Institute of Space Sciences, Shandong University,  \\
 Weihai, Shandong, 264209, China.}

\author{Ya-Wen Cai}
\affiliation{SDU-ADU Joint Science College, Shandong University, Weihai, Shandong, 264209, China.}

\author{Ya-Ni Guo}
\affiliation{Shandong Key Laboratory of Optical Astronomy and Solar-Terrestrial Environment,\\
 School of Space Science and Physics, \\
 Institute of Space Sciences, Shandong University,  \\
 Weihai, Shandong, 264209, China.}

\author{Xing Gao}
\affiliation{Xinjiang Astronomical Observatory, 150 Science 1-Street, Urumqi 830011, China.}

\author{Xi Wang}
\affiliation{Shandong Key Laboratory of Optical Astronomy and Solar-Terrestrial Environment,\\
 School of Space Science and Physics, \\
 Institute of Space Sciences, Shandong University,  \\
 Weihai, Shandong, 264209, China.}

\author{Shi-Peng Yin}
\affiliation{Shandong Key Laboratory of Optical Astronomy and Solar-Terrestrial Environment,\\
 School of Space Science and Physics, \\
 Institute of Space Sciences, Shandong University,  \\
 Weihai, Shandong, 264209, China.}

\author{Fei Liu}
\affiliation{Shandong Key Laboratory of Optical Astronomy and Solar-Terrestrial Environment,\\
 School of Space Science and Physics, \\
 Institute of Space Sciences, Shandong University,  \\
 Weihai, Shandong, 264209, China.}

\author{Guo-You Sun}
\affiliation{Wenzhou Astronomical Association, Beibaixiang Town, Yueqing, Zhejiang, 325603, China.}

\begin{abstract}
The photometric analysis and spectroscopic study of the long period low mass ratio deep contact binary KN Per were executed. The light curves of BV(RI)$_c$-band were from the Ningbo Bureau of Education and Xinjiang Observatory Telescope (NEXT) at the Xingming Observatory. Through the analysis of Wilson-Devinney (W-D) program, KN Per was found as an A-type low mass ratio deep contact binary (q=0.236, f=53.4\%). A cool spot applied on the primary component was introduced to explain the unequal maxima of the light curve. Based on the O-C analysis, we found that the rate of the increasing orbital period is $\dot{P}$ = 5.12 $\pm$ (0.30) $\times$ 10$^{-7}$ d/yr, meaning the mass transfer from the secondary component to the primary one. By analyzing the spectroscopic data, we detected chromospheric activity emission line indicators, which is corresponding to the light-curve analysis. 71 long period (P $>$ 0.5 days) contact binaries including our target were collected. The evolutionary states of all collected stars were investigated by the illustrations of mass-radius, mass-luminosity, and log M$_{T}$ - log J$_{o}$. The relations of some physical parameters were also determined. With the instability parameters of KN Per, we determined that it is a stable contact binary system at present.
\end{abstract}


\section{Introduction}

In the Galactic Disc, an abundant part (1 in 500) was considered to be contact binaries \citep{Rucinski2006}. The contact binaries have two components, whose spectral types are from F to K \citep{Rucinski1993}. A common convective envelope is existing in the contact binaries, which is shared by the primary and secondary stars \citep{Lucy1968}. The contact binaries are divided into two types \citep{1970VA.....12..217B}: A-type and W-type. For A-type binary, the surface temperature of the more massive star is higher than that of the less massive star, while W-type binary is quite the reverse. For contact binaries, the orbital periods are usually less than 0.7 days \citep{Hilditch2001}.

Deep, low mass ratio overcontact binaries (DLMROBs) are a special kind of contact binaries, whose mass ratio are less than 0.25 and contact degree are more than 50\% \citep{2005AJ....130..224Q}. They are thought to be at the late status in the evolution of close binary systems and may merge into fast rotating stars \citep{2008AJ....136.1940Q}. Thus, DLMROBs may be the progenitors of blue straggler \citep{2015AJ....150...69Y}, FK Com-type stars \citep{2001ApJ...562.1012E, 2012JASS...29..145E} and luminous red novas \citep{2011A&A...528A.144K, 2011A&A...531A..18S, 2011A&A...528A.114T, 2016RAA....16...68Z}. Therefore, their photometric and spectrometric investigations are important.

Numerous contact binaries show chromospheric activity, because chromospheric activity usually exists in the kind of systems with spectral type later than F \citep{2008MNRAS.389.1722E}. Magnetic activities, which usually manifest themselves as star spots, flares, and plages \citep{2003ChJAS...3..361P}, have a relationship with convective motions and rapid stellar rotation \citep{2009A&ARv..17..251S}. For the photometric light curves, contact binaries will exhibit the O'Connell effect, the meaning of which is the two unequal maxima of the light curve \citep{1951PRCO....2...85O}. For the spectroscopic data, there are special emission lines indicating chromospheric activity, such as Ca II H and K, Balmer series and Ca II triplet above the continuum \citep{2018A&A...615A.120P, 2019MNRAS.487.5520L, 2020MNRAS.495.1252Z, 2021MNRAS.506.4251Z}.

KN Per was determined as a RR Lyrae star by \cite{1976MmSAI..47..229P}. They also derived the period of this target as 0.433224 days. A new period was determined by \cite{1979ApJS...41..413E}, which is 0.232 days. Based on this period, \cite{1982AJ.....87.1395K} proposed that KN Per is a $\delta$ Scuti star. The period was updated to be 0.433 days by \cite{1985IBVS.2681....1K} and renewed as 0.866448 days by \cite{1991AJ....102.1766S} based on the light curves in VR$_{c}$ band. \cite{1993AJ....106.2429S} also showed light curves in two bands (VR$_{c}$). Then, this target was classified into a faint $\delta$ Scuti star by \cite{1997A&A...327..240A}. Finally, the first photometric analysis of light curves in VR$_{c}$ band was carried out by \cite{1997Ap&SS.254..295G} and a cool spot was applied on the secondary star. They found that KN Per is a DLMROB (q=0.25, f=54.49\%) with a period of 0.8664606 days. Recently, a O-C analysis was reported by \cite{2018NewA...62...20H}. They found that the orbital period shows a continuous increase. In this paper, new photometric and spectroscopic investigations of KN Per are presented.

\section{Observations}
\subsection{Photometric Observations}
The photometric observations of KN Per are from NEXT at Xingming Observation and other available light-curve databases, Transiting Exoplanet Survey Satellite (TESS) \citep{2015JATIS...1a4003R}, All-Sky Automated Survey for Supernovae (ASAS-SN) \citep{2014ApJ...788...48S, 2019MNRAS.486.1907J} and Wide Angle Search for Planets (SuperWASP) \citep{2010A&A...520L..10B}.

The observations of NEXT at Xingming Observation were carried out on November 14, 16, 17, 19 and 20, 2019. The aperture of NEXT is 60 cm. A CCD camera with 2048$\times$2048 pixels is included in NEXT, resulting a field of view around $22' \times 22'$. We adopted the standard Johnson-Cousin-Bessel UBV(RI)$_c$ filter system for the observation. All raw images were corrected for bias, dark and flat using the C-Munipack\footnote[1]{http://c-munipack.sourceforge.net/} program. C-Munipack is a software to process CCD photometric data. The CCD data were reduced using aperture photometry and differential photometry methods. Additionally, the magnitude differences among KN Per, the comparison and the check stars were obtained. Table 1 shows the photometric observation details of KN Per. The exposure times in B-band, V-band, R$_{c}$-band and I$_{c}$-band are 20s, 12s, 8s and 12s, respectively. The observing errors in BV(RI)$_c$-band are shown in Table 2. We display the photometric data in Table 3 and show the light curves in Figure 1. The orbital phases can be determined by the following equation, using the period of \cite{1997Ap&SS.254..295G},
\begin{equation}
\ T = T_{0} + 0.8664606 \times E, \\
\end{equation}
where T is the observed time, T$_{0}$ is a primary minimum obtained by the light curves and E is the epoch for corresponding T. For NEXT data, T are in HJD and T$_{0}$ is 2458802.11355. Except for NEXT, KN Per was also observed by TESS, ASAS-SN and SuperWASP. Sector 18 of TESS data has measured KN Per in a 30-minute cadence from November 4, 2019 to November 26, 2019. ASAS-SN has measured KN Per with the exposure time of 90s from January 15, 2015 to September 21, 2018, and SuperWASP has measured this target with the exposure time of 30s from August 13, 2004 to December 24, 2007. For TESS data, T in Equation (1) is in BJD and T$_{0}$ was 2458792.58670. For ASAS-SN and SuperWASP, we used different ephemerides. T in Equation (1) is in HJD and T$_{0}$ were 2458424.33330 and 2454022.65245, respectively. The deeper primary minimum was always chosen as phase zero. The three light curves of TESS, ASAS-SN and SuperWASP are shown in Figure 2.
\subsection{Spectroscopic Observations}
Having an aperture of 4 m, the Large Sky Area Multi-Object Fiber Spectroscopic Telescope (LAMOST) is a particular reflecting Schmidt telescope with a field of view of $5^{\circ}$ and collects numerous stellar spectra \citep{2012RAA....12.1243L}. The range of the wavelength covered by LAMOST is from 3700{\AA} to 9000 {\AA} \citep{2012RAA....12.1197C, 2016ApJS..227...27D}. The resolution is almost 1800 in the state of low resolution mode \citep{1996ApOpt..35.5155W}. 4000 fibers, located on the surface, improve the rate to acquire spectra greatly  \citep{2012RAA....12.1197C}. For low resolution mode, two spectra of KN Per in 2012 and 2016 were found from Data Release 8\footnote[2]{http://www.lamost.org/dr8/}. We display the spectral parameters in Table 4.

\begin{figure}[h]
\centering
\includegraphics[width=14cm]{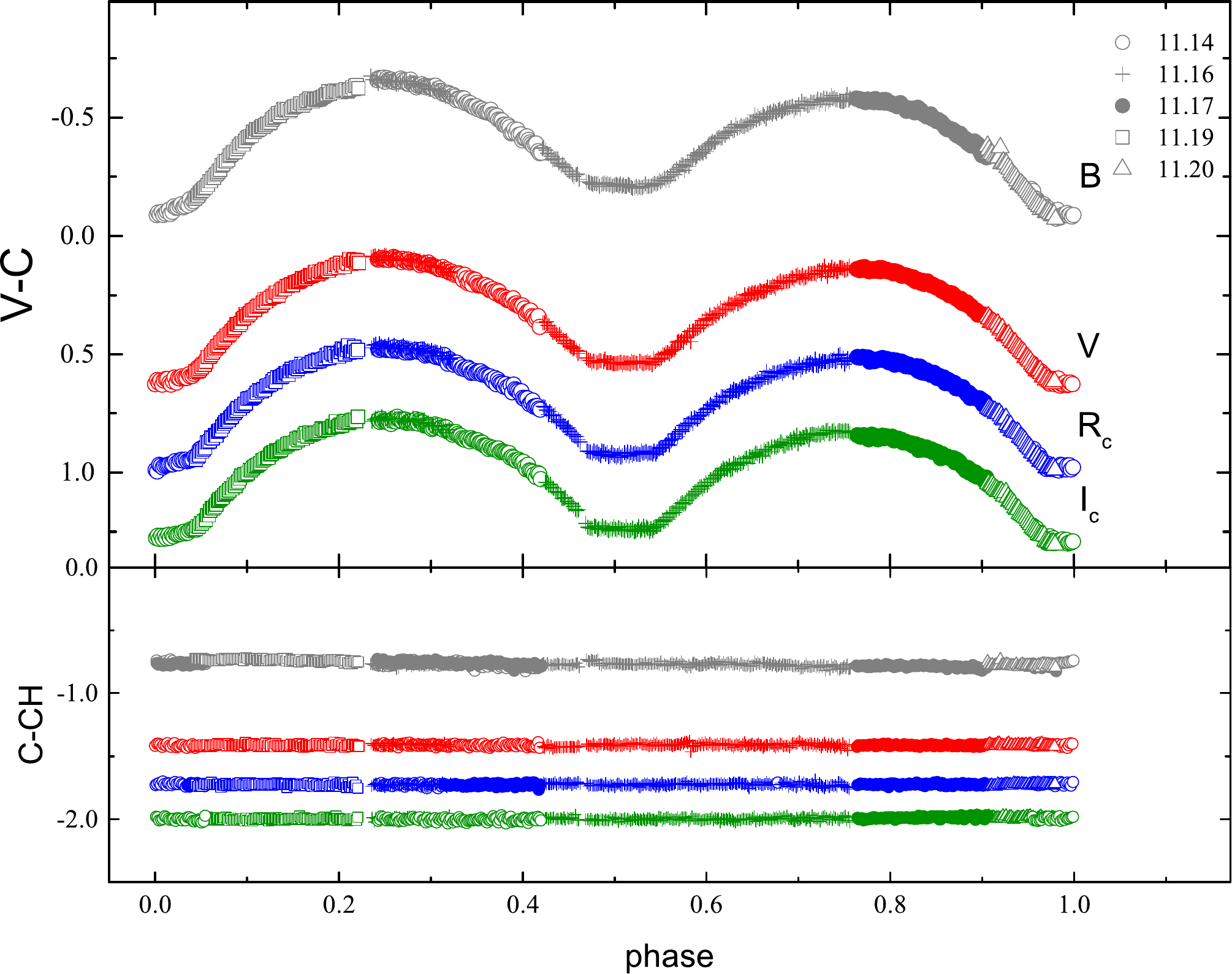}
\caption{Four-band light curves of KN Per. The gray, red, blue and green symbols represent the light curves in B-band, V-band, R$_{c}$-band, I$_{c}$-band. The open circles, crosses, solid circles, hollow squares and hollow triangles represent the data on November 14, 16, 17, 19 and 20, 2019. }
\end{figure}

\renewcommand\arraystretch{1.3}
\begin{table*}
\small
\begin{center}
\caption{Observing details of KN Per}
\begin{tabular}{p{2.5cm}p{4.5cm}p{2cm}p{2cm}p{0.7cm}p{0.7cm}p{0.7cm}}
\hline
Stars            & Name            & $\alpha_{2000}$  & $\delta _{2000}$    & J    & H     &K         \\
 &                 & h m s     &    $^{\circ}$ ' "&mag  &mag    &mag  \\
\hline
KN Per       & KN Per            & 03 22 35.853      & +41 19 55.044         & 10.520 & 10.352  & 10.283   \\
The comparison & 2MASS 03231632+4120289          & 03 23 16.546 & +41 20 29.143 & 9.047 & 8.388 & 8.279   \\
The check      & 2MASS 03220492+4114261 & 03 22 05.145 & +41 14 26.419 & 11.576 & 11.342  & 11.264 \\
\hline
\end{tabular}
\end{center}
\end{table*}

\renewcommand\arraystretch{1.3}
\begin{table*}
\normalsize
\begin{center}
\caption{Observing errors of KN Per}
\begin{tabular}{p{1cm}p{2.1cm}p{2.1cm}p{2.1cm}p{2.2cm}p{2.2cm}p{2.2cm}p{2.2cm}}
\hline
Band           & Nov. 14            &Nov. 16 &Nov. 17    &Nov. 19   &Nov. 20    \\
         &mag           &mag &mag    &mag   &mag    \\
\hline
B        & 0.017  & 0.014    & 0.018        & 0.010& 0.011  \\
V   &    0.010      &  0.012          &   0.006          & 0.008 & 0.017  \\
$R_c$  &    0.011      &    0.011      &  0.007    &   0.009  & 0.006   \\
$I_c$    &   0.013       &  0.012     &   0.010       &  0.008      &  0.007  \\
\hline
\end{tabular}
\end{center}
\end{table*}

\renewcommand\arraystretch{1.3}
\begin{table*}
\normalsize
\begin{center}
\caption{The photometric data of KN Per obtained on November 14, 16, 17, 19 and 20, 2019}
\begin{tabular}{cccccccc}
\hline
HJD$_{B}$ &$\Delta$m$_{B}$ & HJD$_{V}$ &$\Delta$m$_{V}$ & HJD$_{R_{c}}$ &$\Delta$m$_{R_{c}}$ & HJD$_{I_{c}}$ &$\Delta$m$_{I_{c}}$ \\
2400000+ &mag       &2400000+     &mag        &2400000+    &mag        & 2400000+    &mag \\
\hline
58802.07325  &   -0.198  &   58802.07369    &   0.538  &   58802.07405   &   0.900     &    58802.07437    &   1.224   \\
58802.07486  &   -0.188  &   58802.07529    &   0.558  &   58802.07567   &   0.901     &    58802.07600    &   1.229   \\
58802.07647  &   -0.159  &   58802.07692    &   0.567  &   58802.07729   &   0.923     &   58802.07764    &   1.233   \\
58802.07812  &   -0.154  &   58802.07860    &   0.563  &   58802.07898   &   0.923     &    58802.07933    &   1.251   \\
58802.07984  &   -0.145  &   58802.08026    &   0.567  &   58802.08066   &   0.926     &    58802.08100    &   1.257   \\
...          &...&...&...&...&...&...&...\\
58808.15838  &   -0.085   &  58808.15729   &    0.617 &   58808.15763   &   0.980  &  58808.15793   &  1.296 \\
58808.15986  &   -0.089   &  58808.15878   &    0.623 &   58808.15912   &   0.977  &  58808.15941   &  1.296 \\
58808.16133  &   -0.071   &  58808.16027   &    0.615 &   58808.16060   &   0.988  &  58808.16089   &  1.304 \\
\hline
\end{tabular}
\end{center}
This table is available in its entirety in machine-readable form. A portion is shown here for guidance regarding its
form and content.
\end{table*}

\renewcommand\arraystretch{1.3}
\begin{table}
\small
\caption{The Spectroscopic information of KN Per}
\begin{tabular}{p{1.75cm}p{1.95cm}p{1.0cm}p{1.5cm}p{1cm}p{1cm}p{0.8cm}p{2.1cm}p{2.15cm}}
\hline
Data        & HJD           & Phase   & Exposures & Subclass & $T_{eff}$ & log(g) & Radial velocity & H$\alpha$ (6564.61{\AA}) \\
(d)         & (d)           &         & (s)       &          & (K)       &        & $(km s^{-1})$  & \\
\hline
2012 Dec 25 & 2456289.11810 & 0.24393 & 1200      & F0       & 7153.92   & 3.890  & -64.92      & 0.195(0.010)        \\
2016 Nov 16 &  2457709.12542 & 0.07332 & 1800      & F0       & 7129.66   & 3.895  & -24.12        & 0.153(0.010)         \\
\hline
\end{tabular}
\end{table}

\section{Light-curve analysis}

The effective temperature of the primary star, 7142 K, was determined by calculating the average value of the two temperatures shown in Table 4. The average temperature from LAMOST is the temperature of the system T$_{eff}$ and not of the primary. This is usually assigned to the primary star to begin the modelling procedure i.e T$_{1}$$\sim$T$_{eff}$. To ensure its reliability, the temperature obtained by the color indices was used to check it. We calculated the color indices B - V = 0.412 mag, g - r = 0.293 mag and J - K = 0.237 mag from AAVSO Photometric All Sky Survey (APASS) \citep{2016yCat.2336....0H} and Two Micron All-Sky Survey (2MASS)\citep{2003yCat.2246....0C}. E(B - V) = 0.130 mag was obtained from a 3D Dust Map \citep{2019ApJ...887...93G} based on the distance 931.64 (0.02) pc from Gaia EDR3 \citep{2021A&A...649A...1G}. Finally, (B - V)$_0$ = 0.282 mag, (g - r)$_0$ = 0.176 mag and (J - K)$_0$ = 0.173 mag were determined. According to Table 5\footnote[3]{http://www.pas.rochester.edu/\url{~}emamajek/EEM\url{_}dwarf\url{_}UBVIJHK\url{_}colors\url{_}Teff.txt} of \cite{2013ApJS..208....9P}, we obtained the corresponding temperatures (7200 K, 7000 K and 7100 K) through the color indices. Then, the average value of them, 7100 K, was calculated. The difference between this result and the effective temperature obtained by LAMOST data is obviously small. Thus, the approximate value of LAMOST temperature 7140 K ($\pm$ 20 K) was used in the following analysis.

Because the light curves determined from NEXT have four bands and a high photometric accuracy, NEXT data were used to determine the physical parameters by W-D program \citep{1971ApJ...166..605W, 1979ApJ...234.1054W, 1990ApJ...356..613W, 1994PASP..106..921W}. We derived q through the q-search method. The range of q is from 0.17 to 5.0. We set the step size as 0.01 in the range of 0.17$<$q$<$0.40, and set the step size as 0.1 in the range of 0.4$<$q$<$5.0. The following parameters are adjustable: the orbital inclination (i); the effective temperature of the less massive star (T$_{2}$); the luminosity of the more massive star (L$_{1}$) and the potential ($\Omega$ = $\Omega$$_{1}$ =$\Omega$$ _{2}$ ). We show the relation between q and mean residual in Figure 2. The minima of mean residual was seen at q = 0.24. Mean residual = $\sqrt{\frac{\sum_{i} W_{i}(O-C)^{2}_{i}}{N}}$ (N is the total number of all data).

Defining q=0.24 as the initial value and setting q adjustable, the analysis of the NEXT light curves was accomplished by W-D program. The theoretical light curves without spot are shown in Figure 3, which are represented by black lines. We found that the incompatibilities between the theoretical light curves and the observed light curves are obvious. Thus, we applied a cool spot on the primary star to further model the observed light curves. In this fitting process, the value of latitude was fixed at 90 $^{\circ}$. The theoretical light curves of spot model are shown in Figure 3 with red lines, which obviously fit the observed light curves very well. We should declare that the spotted models are not unique.

When we determined the temperature of the primary star, we ignored the existence of the surface brightness of the secondary star, which cannot be negligible. Thus, we assumed blackbody radiation to deal with this problem. 7140 K and 6860 K were assigned as the primary and secondary temperatures. We used the following equations to obtain the temperatures of both primary and secondary stars,
\begin{equation}
T_{1} = (((1 + k^{2})T_{eff}^{4})/(1 + k^{2}(T_{2}'/T_{1}')^{4}))^{0.25},
\end{equation}
\begin{equation}
T_{2} = T_{1}(T_{2}'/T_{1}'),
\end{equation}
where T$_{eff}$ is the effective temperature determined by LAMOST, k is the radius ratio and T$_{2}$'/$T_{1}$' is the temperature ratio \citep{2003A&A...404..333Z, 2013AJ....146..157C}. Then, the primary and secondary temperatures are 7200 K and 6917 K. The errors obtained by W-D program were too small to be true. To determine the more accurate physical parameters and their errors of KN Per, the Marcov Chain Monte Carlo (MCMC) parameter search method by using PHOEBE \citep{2005ApJ...628..426P, 2016ApJS..227...29P, 2018ApJS..237...26H, 2020ApJS..250...34C, 2020ApJS..247...63J} was employed to analyze the NEXT data. The physical parameters obtained by W-D program were set as priors. We set walker as 24. The MCMC parameter search was run for 3000 steps for NEXT light curves and we deleted the first 1000 steps. The probability distributions of q, i, T$_{2}$ and L$_1$/(L$_1$ + L$_2$) are shown in Figure 4. All the physical parameters are displayed in Table 5. The light curves of TESS, ASAS-SN and SuperWASP were analyzed by W-D program, respectively. We set the physical parameters determined from the NEXT light curves as the initial values. The value of latitude was still fixed at 90$^{\circ}$. The theoretical light curves without spot and with a cool spot are all shown in Figure 3. Due to the high accuracy of the light curves observed by NEXT, the result of NEXT was chosen as the final result. The errors of the photometric solutions in Table 5 are formal and underestimated.

\begin{figure}[h]
\centering
\includegraphics[width=14cm]{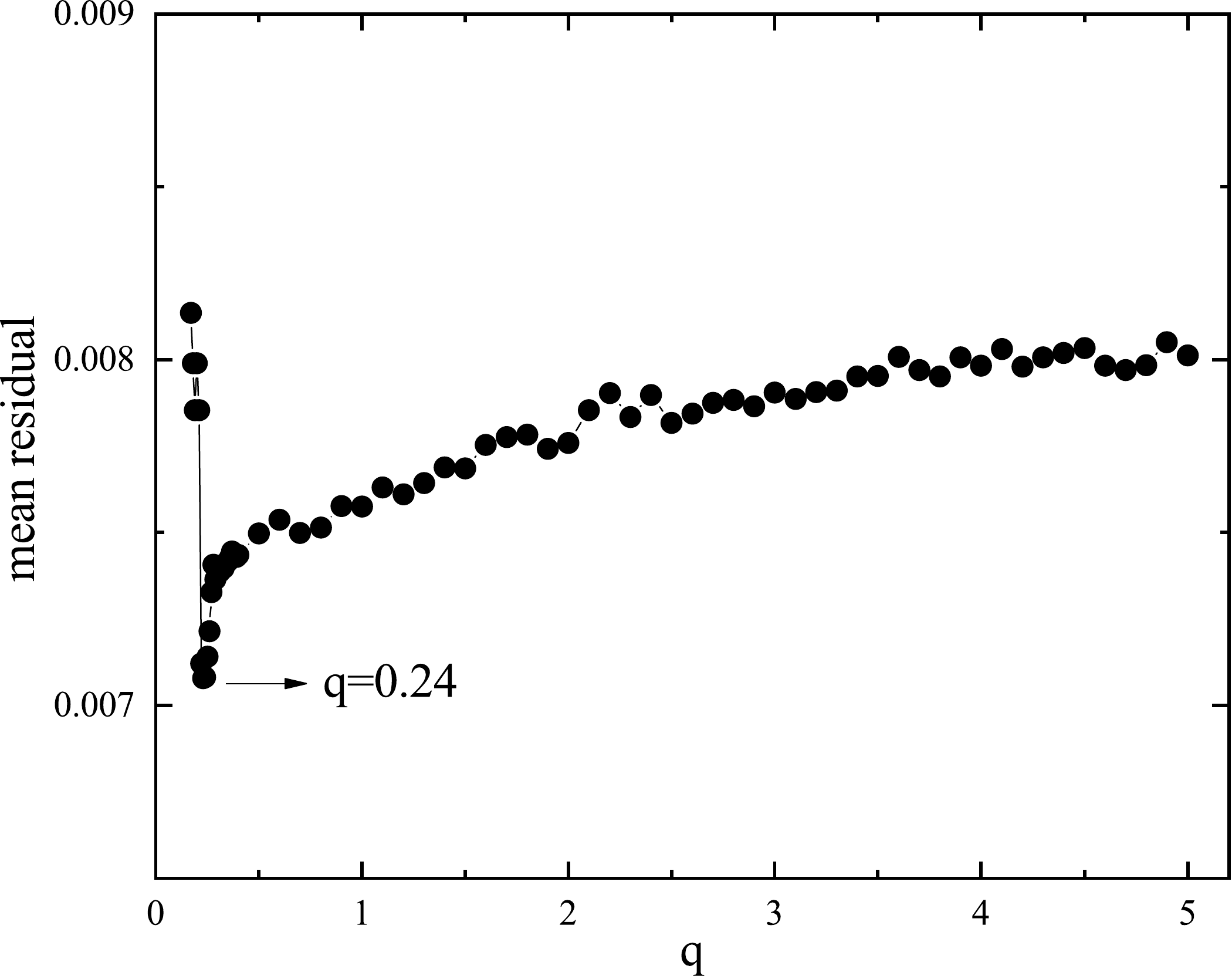}
\caption{The relation between q and mean residual.}
\end{figure}

\begin{figure}
\begin{center}
\includegraphics[width=0.45\textwidth]{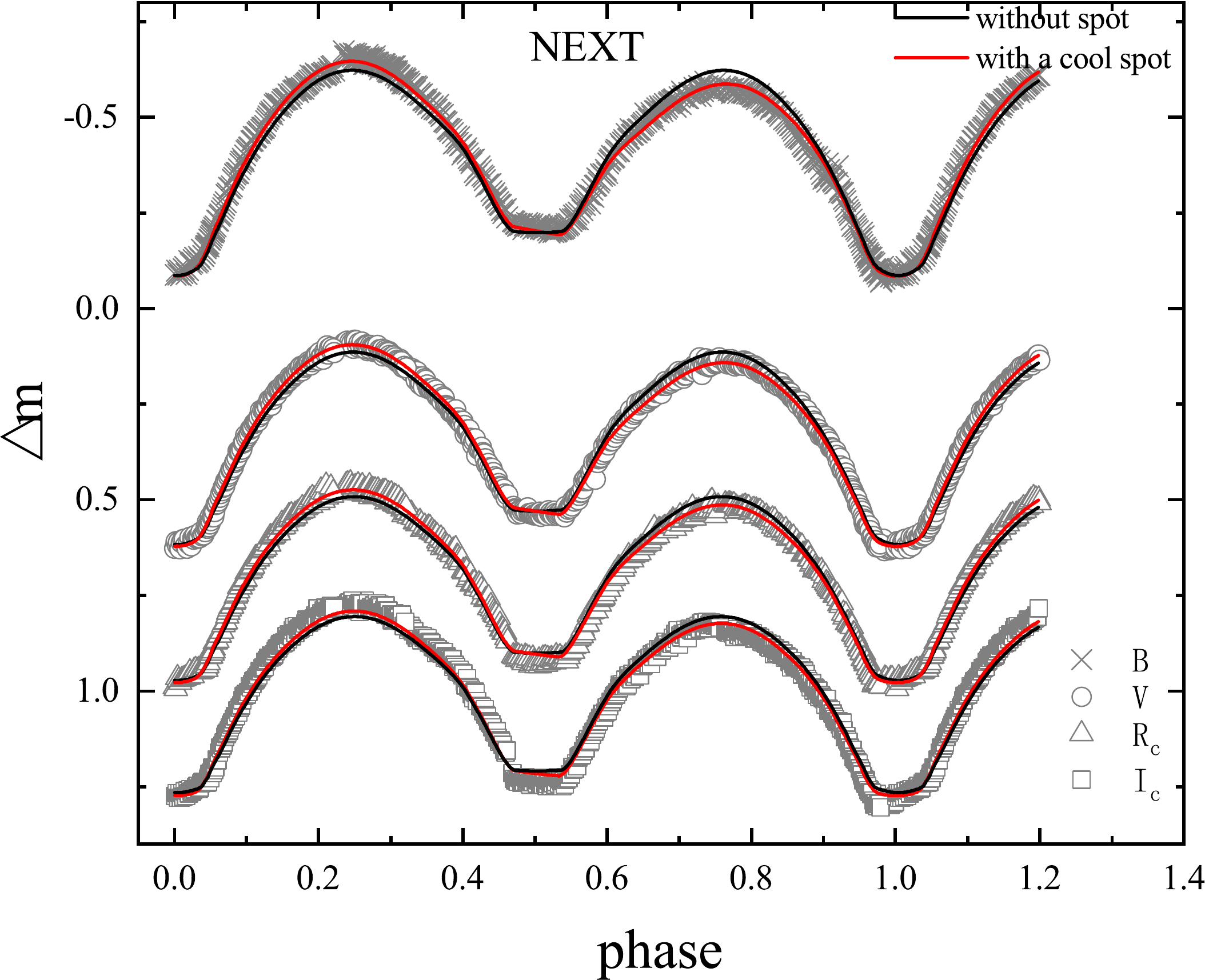}
\includegraphics[width=0.45\textwidth]{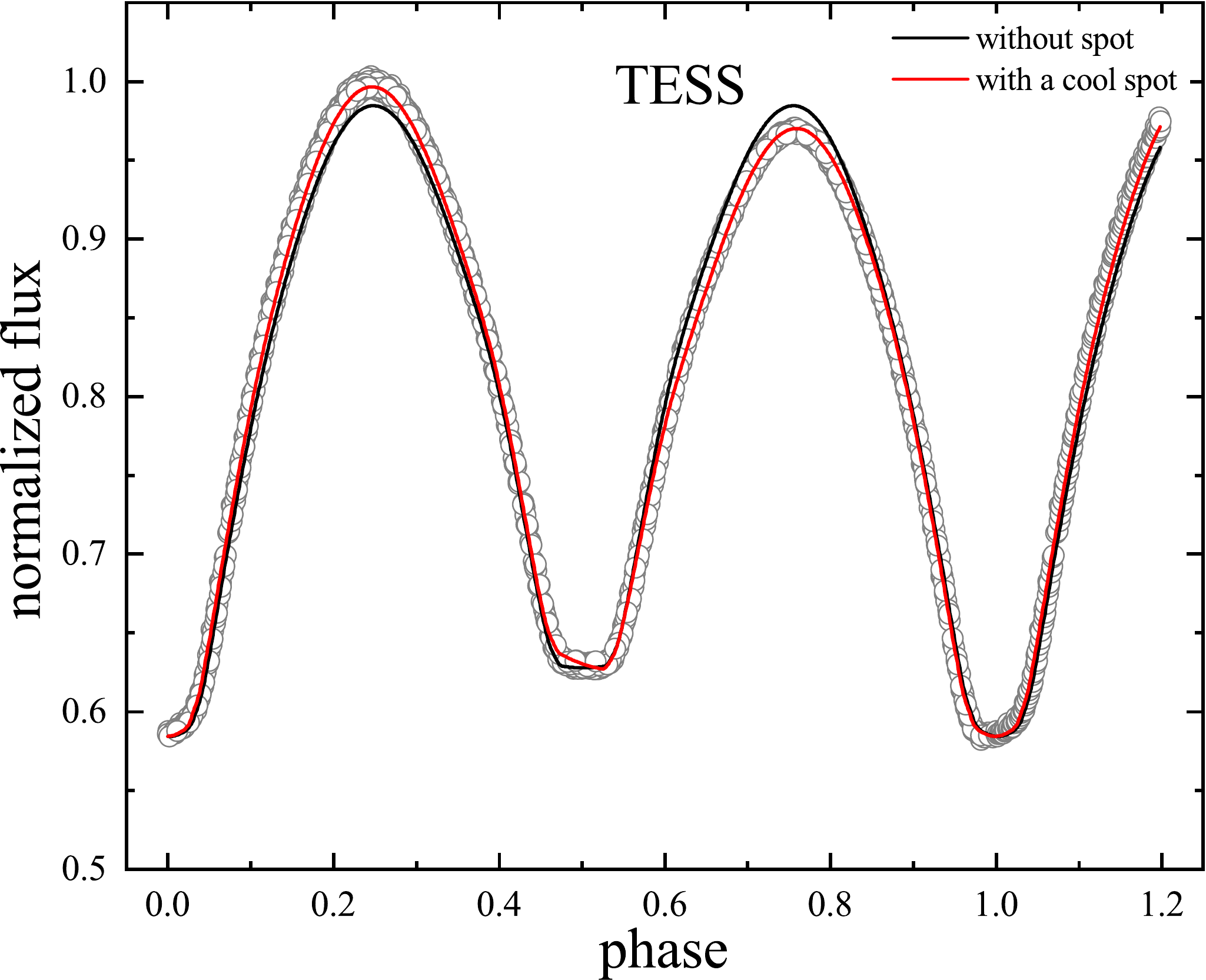}
\includegraphics[width=0.45\textwidth]{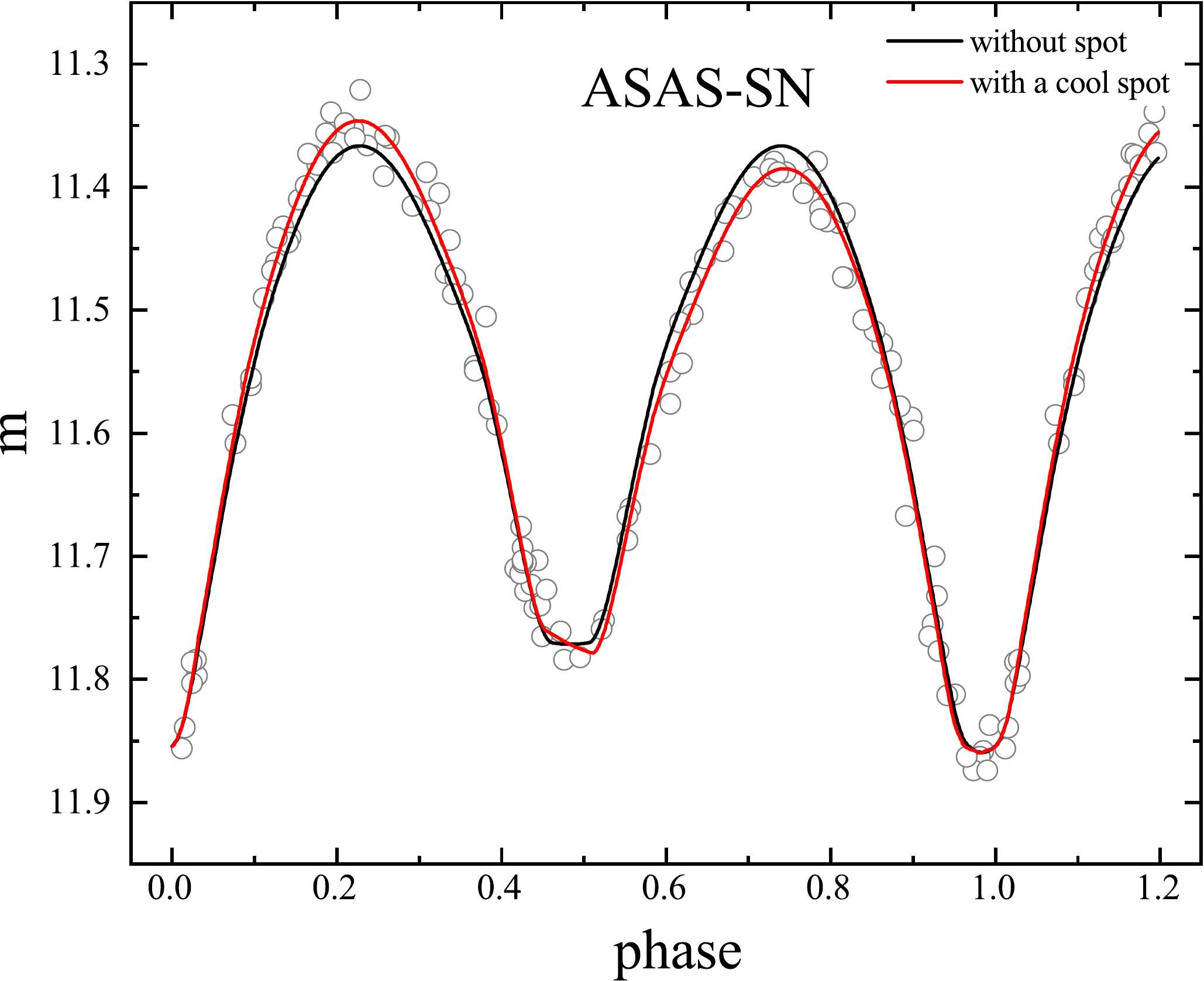}
\includegraphics[width=0.45\textwidth]{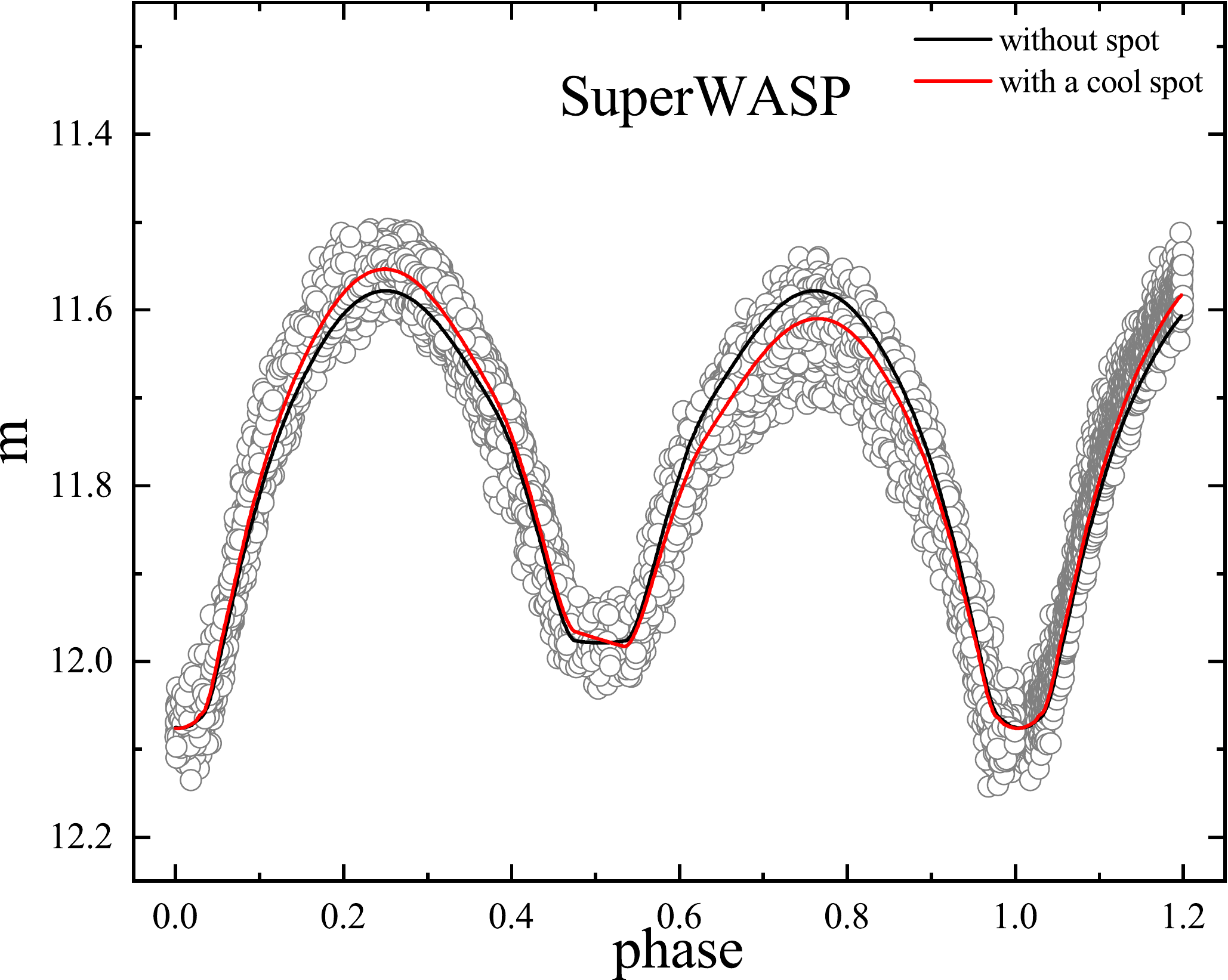}
\caption{Theoretical and observed light curves of KN Per determined by NEXT, TESS, ASAS-SN and SuperWASP. The black lines represent the theoretical light curves without spot and the red lines represent the theoretical light of spot models.}
\end{center}
\end{figure}

\begin{figure}[h]
\centering
\includegraphics[width=14cm]{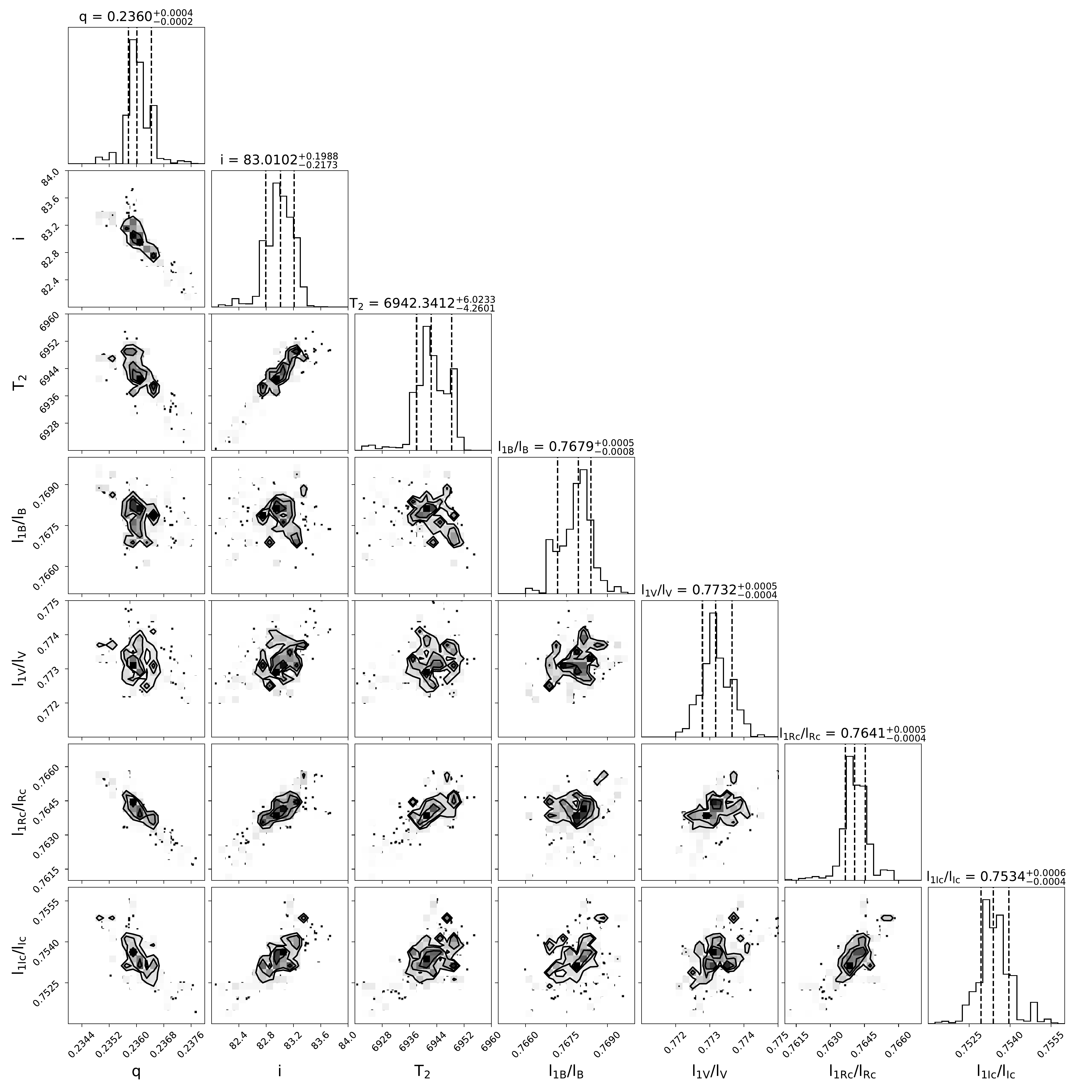}
\caption{The probability distributions of q, i, T$_{2}$ and L$_1$/(L$_1$ + L$_2$) determined by the MCMC modeling for the NEXT light curves.}
\end{figure}

\begin{table}
\centering
\tiny
\caption{Photometric solution of KN Per}
\begin{tabular}{p{6.7cm}p{2.4cm}p{2.4cm}p{2.4cm}p{2.4cm}}
\hline
                &NEXT   &TESS  &ASAS-SN &SuperWASP\\
\hline
Primary temperature $T_1(K)$      &7200        &7200     &7200      &7200  \\
Secondary temperature $T_2(K)$      &6942$^{+6}_{-4}$       & 6875($\pm$4)  & 6824($\pm$40)& 6855($\pm$8) \\
Mass ratio $q(M_2/M_1)$  &0.2360$^{+0.0004}_{-0.0002}$    &0.2430($\pm$0.0007) &0.2356($\pm$0.0088) &0.2498($\pm$0.0016) \\
Orbital inclination i(deg)             &83.01$^{+0.20}_{-0.22}$    &88.93($\pm$0.05) &80.24($\pm$0.08)&81.97($\pm$0.30)\\
Modified dimensionless surface potential of star $\Omega$      &2.259$^{+0.011}_{-0.011}$  &2.242($\pm$0.002)  & 2.217($\pm$0.024)&2.256($\pm$0.005)\\
Luminosity ratio L$_1$/(L$_1$ + L$_2$) in TESS &- &0.7538($\pm$0.0001)&- &-\\
Luminosity ratio L$_1$/(L$_1$ + L$_2$) in SuperWASP &- &-&- &0.8028($\pm$0.0002)\\
Luminosity ratio L$_1$/(L$_1$ + L$_2$) in band B  &0.7679$^{+0.0005}_{-0.0008}$  &- &-&-\\
Luminosity ratio L$_1$/(L$_1$ + L$_2$) in band V  &0.7732$^{+0.0005}_{-0.0004}$ &-&0.8062($\pm$0.0002)&-\\
Luminosity ratio L$_1$/(L$_1$ + L$_2$) in band R$_{c}$   &0.7641$^{+0.0005}_{-0.0004}$   &-&-&-\\
Luminosity ratio L$_1$/(L$_1$ + L$_2$) in band I$_{c}$   &0.7534$^{+0.0006}_{-0.0004}$ &-&- &-\\
Equal-volume radius of star 1 (relative to semimajor axis) $r_1$        &0.5326$^{+0.0003}_{-0.0002}$   &0.5307($\pm$0.0004) &0.5445($\pm$0.0064)&0.5319($\pm$0.0011)\\
Equal-volume radius of star 2 (relative to semimajor axis) $r_2$        &0.2958$^{+0.0060}_{-0.0060}$  &0.3302($\pm$0.0023) &0.3029($\pm$0.0363)&0.2983($\pm$0.0052)\\
Contact degree f($\%$)         &53.4$^{+4.4}_{-5.0}$  &61.9($\pm$1.3)  & 68.4($\pm$16.3) &60.9($\pm$3.0)  \\
Latitude $\theta(^{\circ})$      &90.00   &90.00  &90.00 &90.00  \\
Longitude $\psi(^{\circ})$        &108.59$^{+0.56}_{-0.88}$  &116.01($\pm$1.95)  &118.15($\pm$7.63)  &100.65($\pm$1.71)\\
Spot radius $r(^{\circ})$       &29.36$^{+0.20}_{-0.11}$ &18.13($\pm$0.09)  &17.51($\pm$0.95) &19.39($\pm$0.18) \\
Temperature factor T$_{f}$      &0.893$^{+0.002}_{-0.001}$ &0.843($\pm$0.002) &0.856($\pm$0.022)&0.846($\pm$0.003)\\
\hline
\end{tabular}
\end{table}

\section{O-C diagram}

\begin{figure}[h]
\centering
\includegraphics[width=14cm]{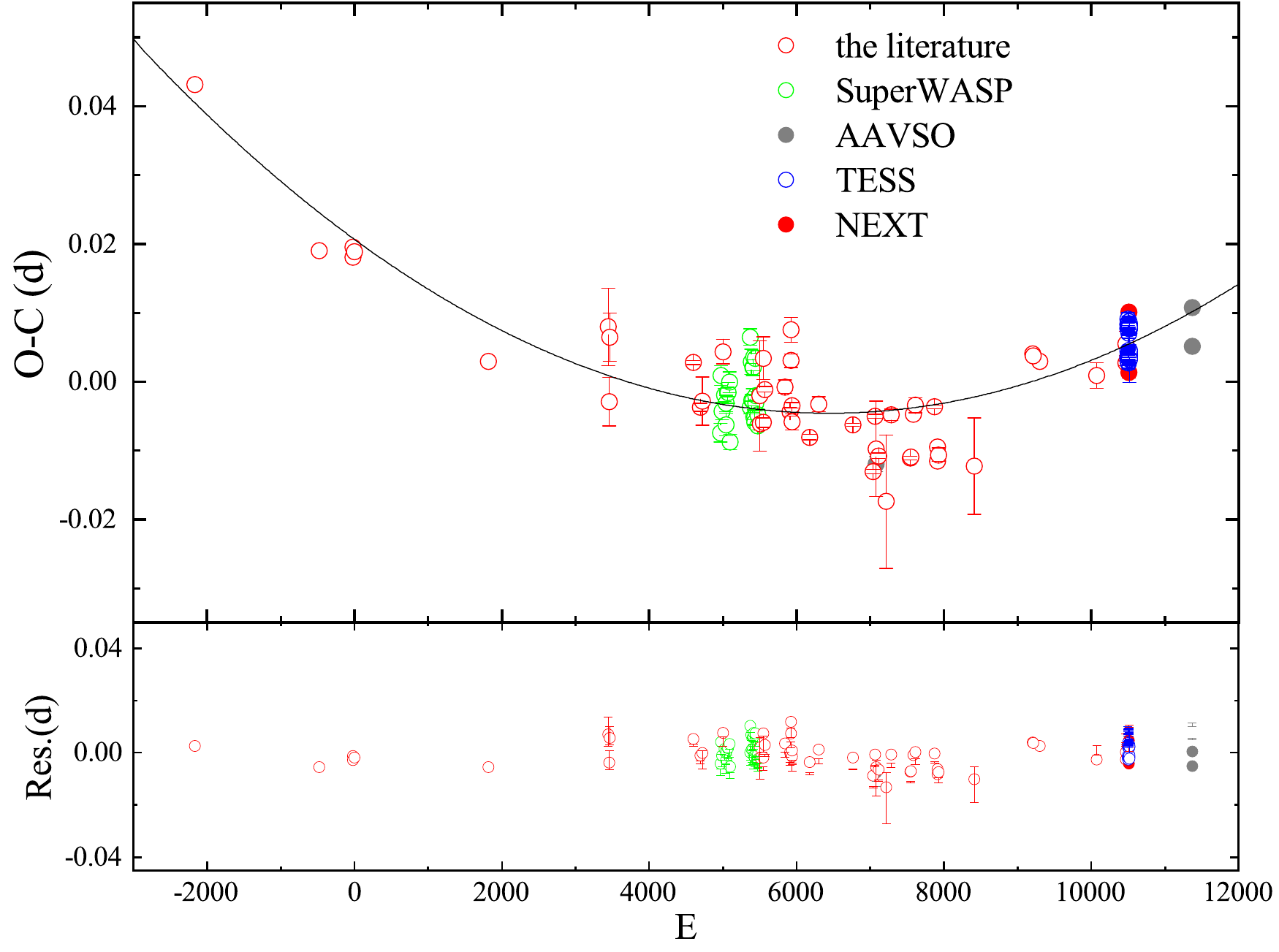}
\caption{The upper illustration refers to O-C diagram. The lower illustration shows the residuals. The data shown in the red hollow points are from the literature, the green hollow points are from SuperWASP, the gray solid points represent data from AAVSO, the blue hollow points are samples obtained from TESS and the red solid points are from NEXT.}
\end{figure}

We derived a total of 121 minimum timings to construct the O-C diagram. Minima calculated by us are shown as follows: 2 minimum timings determined from NEXT, 41 minimum moments from TESS, 3 minima obtained from the American Association of Variable Star Observers (AAVSO) and 26 minima from SuperWASP. In addition, 49 available moments were collected from the literature. We found that the two minimum timings, 2449697.6818 and 2449697.6823, from \cite{1997Ap&SS.254..295G}, are the same. Thus, the average value calculated by us was used in the following analysis. The photographic and visual minimum timings were not included in the following analysis due to their low accuracy. The minima from TESS are in BJD, the minimum timings from AAVSO are in JD, and the remaining are in HJD, which were turned into BJD through Time Utilities\footnote[4]{https://astroutils.astronomy.osu.edu/time/}. Table 6 shows the minimum timings. Based on the initial ephemeris presented by \cite{2003ChJAS...3..361P}: Min.I = 2449698.11390 + 0.8664593 $\times$ E, we calculated the values of O-C. Then, a linear fitting was applied to the O-C values. A new ephemeris was determined: Min.I = 2449698.09710($\pm$0.00205) + 0.8664715($\pm$0.000003) $\times$ E. Finally, new O-C values were obtained by the above ephemeris and the O-C diagram is shown in Figure 5. The data shown in the red hollow points are data from the literature, the green hollow points are data from SuperWASP, the gray solid points represent the data of AAVSO, the blue hollow points are samples obtained from TESS and the red solid points are from NEXT. From this figure, the O-C curve shows an apparently upward parabolic trend. According to the least square method, the following equation was calculated:
\begin{eqnarray}
\label{E2}
Min.I&=& 2449698.11777(\pm0.00158)
     + 0.866464(\pm0.0000005) \times E \\
\nonumber
     &+& 6.07(\pm0.36) \times 10^{-10} \times E^{2}.
\end{eqnarray}
Based on the coefficient of its second order term, we calculated the rate of the increasing orbital period to be 5.12 $\pm$ (0.30) $\times$ 10$^{-7}$ d/yr.

\renewcommand\arraystretch{1.2}
\begin{table*}
\small
\begin{center}
\caption{The minimum timings of KN Per}
\begin{tabular}{p{1.5cm}p{0.8cm}p{1.2cm}p{1.5cm}p{1.5cm}p{0.5cm}|p{1.5cm}p{0.8cm}p{1cm}p{1.5cm}p{1.5cm}p{0.5cm}}
\hline
BJD       &Error &E   &O-C  &Residual  &Ref.  &BJD      &Error  &E &O-C  &Residual &Ref. \\
2400000+  &      &    &     &          &      & 2400000+ &        &   &      &          &      \\
\hline
47820.92965  &  -     &     -2166.5   &  0.04312  &   0.00089    &  (1)               &              56007.30475   &  0.00080  &  7281.5   &  -0.00482   & -0.00007   &  (18)   \\
49283.94271  &  -     &     -478      &  0.01901  &   -0.00655   &  (1)               &              56233.88077   &  0.00030  &  7543     &   -0.01111  &  -0.00667  &   (22)  \\
49679.92072  &  -     &     -21       &  0.01952  &   -0.00214   &  (1)               &              56239.07977   &  -        &  7549     &   -0.01094  &  -0.00650  &   (23)  \\
49680.78572  &  -     &     -20       &  0.01805  &   -0.00360   &  (1)               &              56268.11277   &   -       &  7582.5   &   -0.00473  &  -0.00035  &   (23)  \\
...          &...&...&...&...&...&...&...&...&...&...&...\\
55863.89775  &  0.00040 &   7116    &    -0.01078&    -0.00589  &   (21)     &        58813.38091  &   0.00069 &   10520   &    0.00328 &    -0.00186 &    (29)   \\
55951.40475  &  0.00970 &   7217    &    -0.01740&    -0.01260  &   (17)     &        58813.81877  &   0.00085 &   10520.5 &    0.00791 &    0.00277  &    (29)   \\
56007.30475  &  0.00080 &   7281.5  &    -0.00482&    -0.00007  &   (18)     &        59550.75006  &   0.00032 &   11371   &    0.00515 &    -0.00478 &    (19)   \\
56007.30475  &  0.00080 &   7281.5  &    -0.00482&    -0.00007  &   (18)     &        59552.92182  &   0.00070 &   11373.5 &    0.01074 &    0.00079  &    (19)   \\
\hline
\end{tabular}
\end{center}
(1) \cite{1997Ap&SS.254..295G}; (2) http://var.astro.cz/ocgate/; (3) \cite{2005IBVS.5643....1H}; (4) \cite{2006IBVS.5677....1D}; (5) \cite{2006IBVS.5731....1H};
(6) \cite{2007IBVS.5761....1H}; (7) SuperWASP; (8) \cite{2009IBVS.5874....1H}; (9) \cite{2008IBVS.5830....1J}; (10) VSB; (11) \cite{2009IBVS.5889....1H};
(12) \cite{2010IBVS.5918....1H}; (13) \cite{2009OEJV..107....1B}; (14) \cite{2009IBVS.5871....1D}; (15) \cite{2010IBVS.5941....1H}; (16) \cite{2011IBVS.5960....1D};
(17) \cite{2012IBVS.6026....1H}; (18) \cite{2013OEJV..160....1H}; (19) AAVSO; (20) \cite{2013IBVS.6084....1H}; (21) \cite{2012IBVS.6011....1D}; (22) \cite{2013IBVS.6042....1D}; (23) \cite{2014IBVS.6094....1C}; (24) \cite{2014OEJV..165....1H}; (25) \cite{2014IBVS.6118....1H}; (26) \cite{2015IBVS.6152....1H}; (27) BRNO; (28) BAVJ; (29) TESS; (30) our data.

Note. This table is available in its entirety in machine-readable form. A portion is shown here for guidance regarding its form and content.
\end{table*}

\section{Spectral Analysis}

\begin{figure*}\centering
\includegraphics[width=0.45\textwidth]{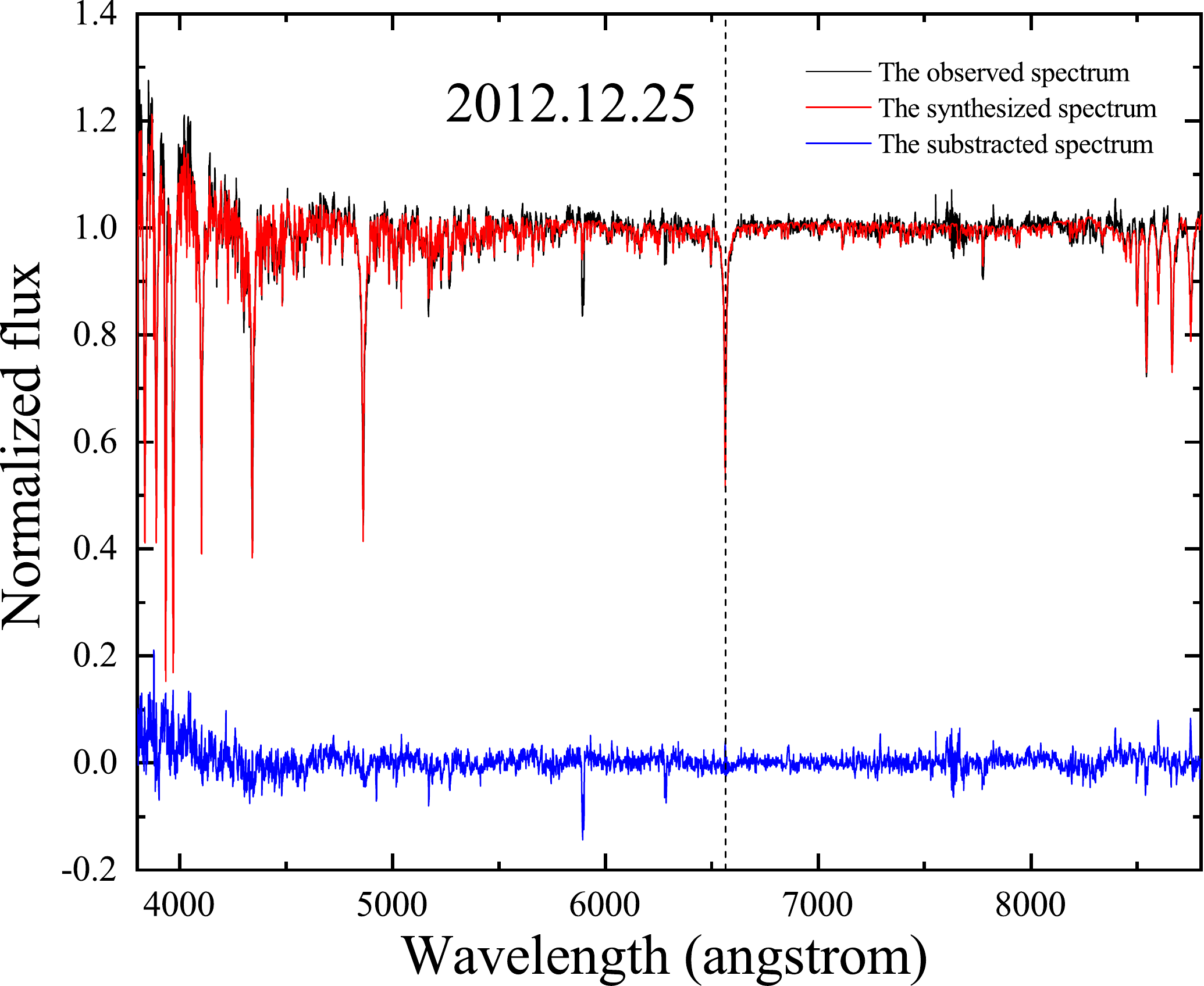}
\includegraphics[width=0.45\textwidth]{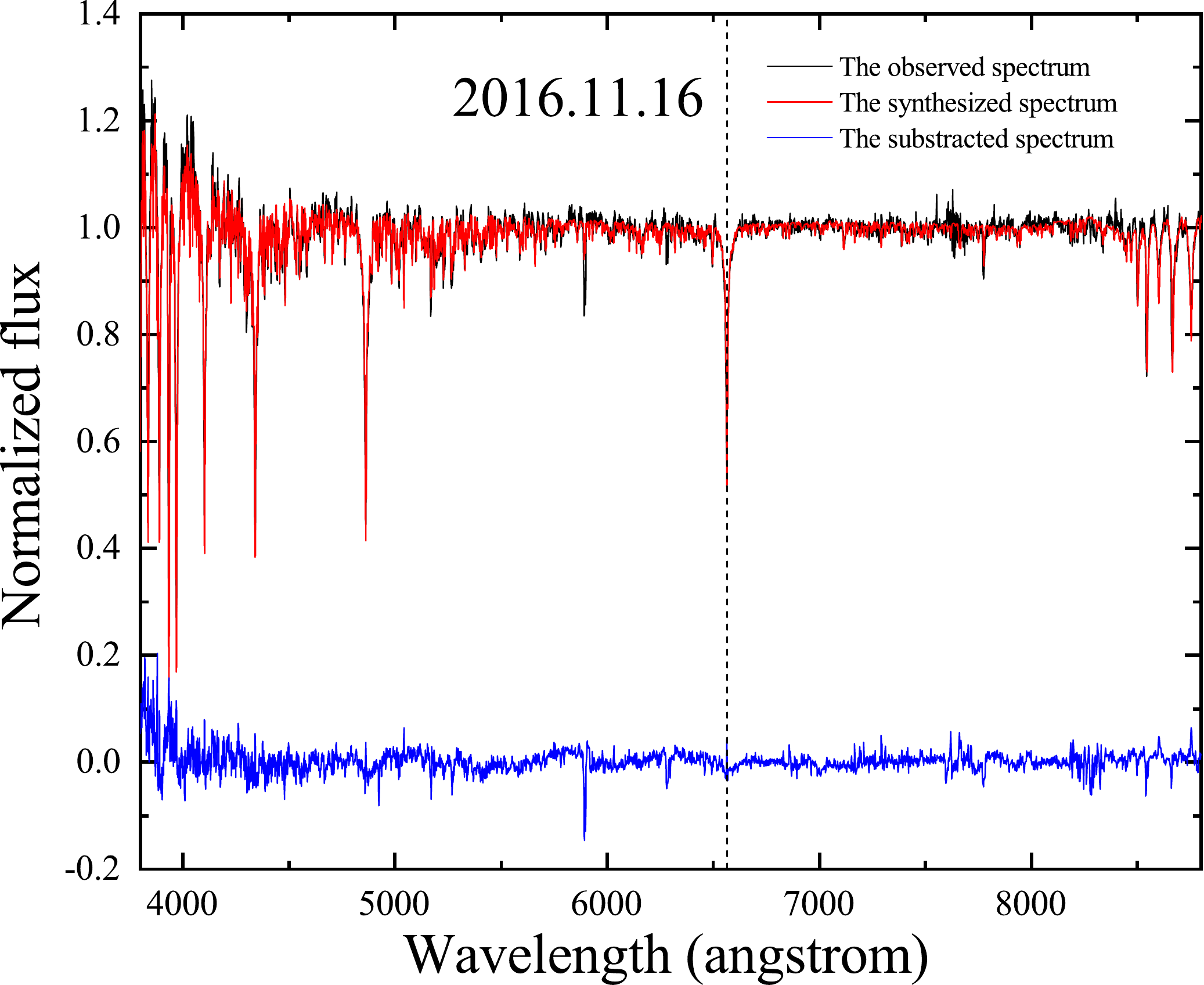}
\caption{The normalized observed, synthesized and subtracted spectra observed by LAMOST. The left panel refers to the data on December 25, 2012, while the right panel represents those on November 16, 2016. The black lines represent the observed spectrum, the red lines represent the synthesized spectrum, and the blue lines represent the subtracted spectrum. The dotted lines represent the locations of H$_\alpha$.}
\end{figure*}

The analysis of chromospheric lines was carried out with the spectral subtraction technique \citep{1985ApJ...295..162B}. There are four steps in the analysis process. Firstly, we used iSpec (iSpec is a tool to treat and analyze stellar spectra) to obtain the synthesized spectrum through some input parameters \citep{2014A&A...569A.111B, 2019MNRAS.486.2075B}. The parameters included the effective temperature, surface gravity, metallicity, resolution and the range of wavelength. Secondly, LAMOST spectra of KN Per and the synthesized spectrum were normalized. Thirdly, we determined the wavelength of H$_\alpha$ ($\lambda$) of the synthesized and observed spectra, calculated their difference ($\Delta$$\lambda$) and obtained $\Delta$$\lambda$/$\lambda$. We shifted the synthesized spectrum based on $\Delta$$\lambda$/$\lambda$. Finally, we obtained the subtracted spectra, the difference between the observed spectra and the synthesized spectrum. Figure 6 displays the observed, synthesized and subtracted spectra. The left panel refers to data on December 25, 2012, while the right panel represents those on November 16, 2016. With Splot package of IRAF, the equivalent widths (EWs) of H$_\alpha$ were determined, shown in Table 4.

\section{Discussion and Conclusion}

To further investigate the structure and evolution of KN Per, the photometric and spectroscopic studies were carried out in this paper. The flat bottom near phase 0.5 indicates that KN Per is a totally eclipsing binary. Thus, the photometric solutions are reliable \citep{2003CoSka..33...38P, 2005Ap&SS.296..221T, 2021AJ....162...13L}. Therefore, KN Per is an A-type long period DLMROB (P=0.8664606 days, q=0.236, f=53.4\%). A cool spot on the primary star was used to fit the observed light curves better. For the photometric analysis, the physical parameters obtained by us are similar to those from \cite{1997Ap&SS.254..295G}. For the spectroscopic analysis, the magnetic activity emission line indicator, H$\alpha$, existing in the subtracted spectra, means that KN Per has weak magnetic activity. With all available minimum times, an upward parabolic trend are displayed in the O-C diagram. The increasing rate of period is 5.12 $\pm$ (0.30) $\times$ 10$^{-7}$ d/yr, which is due to the mass transfer from the secondary component to the primary component. The material transfer rate is calculated as 1.28 $\times$ 10$^{-7}$ M$_\odot$/yr.

\renewcommand\arraystretch{1.3}
\begin{sidewaystable}
\tiny
\begin{center}
\caption{The long period (P$>$0.5 days) contact binaries with radial velocity observations}
{
\begin{tabular}{p{1.5cm}p{0.5cm}p{0.5cm}p{0.5cm}p{0.4cm}p{0.4cm}p{0.4cm}p{0.4cm}p{0.4cm}p{0.4cm}p{0.2cm}p{0.4cm}p{0.4cm}p{0.4cm}
p{0.5cm}p{0.5cm}p{1cm}p{1cm}p{0.4cm}p{0.4cm}p{0.8cm}p{0.8cm}p{1.5cm}}
\hline
Star& Type & V$_{max}$&  P &  q$_{ph}$  & q$_{sp}$& r$_{1}$& r$_{2}$& T$_{1}$& T$_{2}$&  f &  i &  a&  M$_{1}$	&  M$_{2}$&L$_{1}$	 & L$_{2}$	&  R$_{1}$	& R$_{2}$  & J$_{s}$/J$_{o}$ & log J$_{o}$  &Ref.\\
 &&(mag)&(days) & &&& &(K)&(K)&(\%) &($^{\circ}$) &(R$_\odot$)&(M$_\odot$) & (M$_\odot$) & (L$_\odot$)& (L$_\odot$)& (R$_\odot$)&(R$_\odot$)& & (cgs)& \\
\hline
OO Aql  &A  & 9.50  &0.5068 & 0.843 & 0.846 & 0.4130 &	0.3840& 6100	&5926 & 37 & 87.7 &3.35 	&  1.06&  0.90& 	2.45 	&    1.89 	&1.41  &1.31  &0.0387 &  51.88  & (1)           \\
BV Eri  &A  & 8.13  &0.5077 & 0.253 & 0.297& 0.5156 &	0.3077 & 6881	&5737 & 37 & 74.5 &2.99 	&  1.07&  0.32& 	4.65 	&    0.77 	&1.51  &0.89  &0.0861 &  51.44  & (2),(3)       \\
XX Sex  &A  & 9.32  &0.5401 & 0.100 & 0.100 & 0.5940 &	0.2210 & 6881	&6378 & 42 & 74.9 &3.15 	&  1.30&  0.13& 	7.02 	&    0.72 	&1.87  &0.70  &0.2361 &  51.18  & (1)           \\
V357 Peg &A & 9.01  &0.5785 & 0.401 & 0.401& 0.5410 &	0.3190 & 7000	&6438 & 10 & 73.2 &3.92 	&  1.72&  0.69& 	9.67 	&    2.41 	&2.12  &1.25  &0.0699 &  51.96  & (4),(5)       \\
V401 Cyg &A & 10.53 &0.5827 & 0.300 & 0.290 & 0.5111 &	0.3070& 6700	&6650 & 46 & 77.0 &3.80	&  1.68&  0.49& 	6.57 	&    2.39 	&1.91  &1.17  &0.0753 &  51.83  & (6),(7)       \\
$\epsilon$ CrA  &A & 4.74  &0.5914 & 0.128 & 0.137 & 0.5688 &	0.2290& 6678	&6341 & 25 & 0.7  &3.69 	&  1.70&  0.23& 	7.75 	&    1.02 	&2.10  &0.85  &0.1746 &  51.50  & (8),(9)       \\
HI Dra  &A  & 9.02  &0.5974 & 0.250 & 0.250 & 0.5130 &	0.2798 & 7220	&6890 & 24 & 54.0 &3.86 &  1.72&  0.43& 	9.60 	&    2.40 	&1.98  &1.08  &0.0848 &  51.78  & (10),(11),(12)\\
V402 Aur &A & 8.84  &0.6035 & 0.201 & 0.201 & 0.5237 &	0.2505 & 6775	&6700 & 3  & 0.5  &3.77 	&  1.64&  0.33& 	7.43 	&    1.49 	&2.00  &0.92  &0.1029 &  51.66  & (13),(14)     \\
RR Cen  &A  & 7.27  &0.6057 & 0.205 & 0.210& 0.5353 &	0.2782 & 6912	&6891 & 35 & 81.0 &3.92 	&  1.82&  0.38& 	8.89 	&    2.20 	&2.10  &1.05  &0.1067 &  51.74  & (15)          \\
BX AND  &A  & 9.05  &0.6101 & 0.623 & 0.445 & 0.4500 &	0.3200& 6650	&4758 & 5  & 75.9 &4.42	&  2.15&  0.98& 	7.08 	&    0.90 	&2.01  &1.40  &0.0416 &  52.22  & (16)          \\
UZ Leo  &A  & 9.61  &0.6180 & 0.233 & 0.303& 0.5140 &	0.3130 & 6980	&6772 & 76 & 87.4 &4.21 	&  2.01&  0.62& 	10.60 &    3.68 	&2.23  &1.40  &0.0911 &  51.91  & (17)          \\
SZ Hor  &A  & 11.01 &0.6251 & 0.470 & 0.473 & 0.4550 &	0.3230& 6881	&6169 & 18 & 80.4 &4.38&  1.82&  0.86& 	7.61 	&    1.23 	&1.95  &1.38  &0.0481 &  52.08  & (2)           \\
V535 Ara &A & 7.17  &0.6293 & 0.361 & 0.302& 0.4976 &	0.2921 & 8200	&8129 & 22 & 79.9 &4.20 	&  1.94&  0.59& 	18.00 &    6.00 	&2.09  &1.23  &0.0630 &  51.99  & (18)          \\
V407 Peg &A & 9.70  &0.6369 & 0.256 & 0.256& 0.4700 &	0.3120 & 6980	&6484 & 81 & 71.1 &4.02 	&  1.72&  0.43& 	9.80 	&    2.33 	&2.15  &1.21  &0.0724 &  51.79  & (4),(17)      \\
V868 Mon &A & 8.90  &0.6377 & 0.373 & 0.373 & 0.4960 &	0.3290& 7000	&6584 & 49 & 72.1 &4.45 	&  2.39&  0.89& 	14.70 &    3.89 	&2.33  &1.56  &0.0633 &  52.18  & (10)          \\
FP Boo   &A & 10.14 &0.6405 & 0.100 & 0.096& 0.5944 &	0.2147& 6980	&6456 & 38 & 68.8 &3.78 	&  1.61&  0.15& 	11.19 &    0.92 	&2.31  &0.77  &0.2362 &  51.36  & (19),(20)     \\
AG Vir  &A & 8.50  &0.6427 & 0.382 & 0.382 & 0.4890 &	0.3240& 8150	&6953 & 17 & 84.4 &4.48 	&  2.18&  0.74& 	19.00 &    3.85 	&2.19  &1.36  &0.0606 &  52.11  & (21)          \\
FN Cam  &W  & 8.58  &0.6771 & 0.222 & 0.222 & 0.5569 &	0.3178 & 6700	&6848 & 88 & 71.2 &4.65 &  2.40&  0.53& 	12.11 &    4.32 	&2.59  &1.48  &0.1098 &  51.99  & (22)          \\
V1073 Cyg &A& 8.52  &0.7859 & 0.319 & 0.303 & 0.4920 &	0.2863 & 7300	&6609 & 12 & 68.4 &5.17 &  1.81&  0.55& 	16.48 &    3.75 	&2.55  &1.48  &0.0665 &  51.94  & (21)          \\
MW Pav &W   & 8.51  &0.7950 & 0.222 & 0.228 & 0.5428 &	0.2894& 6900	&6969 & 60 & 86.4 &4.43 	&  1.51&  0.33& 	11.82 &    3.31 	&2.41  &1.28  &0.1034 &  51.67  & (23)          \\
V376 And &A & 7.02  &0.7987 & 0.305 & 0.305& 0.4890 &	0.2850 & 9000	&7080 & 7  & 62.7 &5.32 &  2.44&  0.74& 	40.00 &    5.00 	&2.60  &1.51  &0.0677 &  52.15  & (24)          \\
V2388 Oph &A & 8.48  &0.8023 & 0.186 & 0.186 & 0.5553 &	0.2781 & 6900	&6349 & 65 & 74.7 &4.68 &  1.80&  0.34& 	13.50 &    2.43 	&2.60  &1.30  &0.1235 &  51.73  & (25)          \\
TY Pup  &W  & 8.62  &0.8192 & 0.184 & 0.246  & 0.5170 &	0.3130 & 6900	&6915 & 84 & 83.6 &4.65 &  1.65&  0.30& 	14.11 &    3.86 	&2.64  &1.37  &0.1102 &  51.67  & (2),(26)      \\
II Uma &W  & 8.17  &0.8252 & 0.172 & 0.186 & 0.5682 &	0.2828 & 6550	&6554 & 87 & 77.8 &4.87 	&  1.94&  0.33& 	12.14 &    2.62 	&2.73  &1.32  &0.1376 &  51.76  &(27)         \\
V921 Her &A & 9.44  &0.8774 & 0.227 & 0.226& 0.5113 &	0.2577 & 7780	&7346 & 23 & 68.1 &5.29 &  2.07&  0.51& 	5.09 	&    5.60 	&1.41  &0.25  &0.0898 &  51.94  & (20),(28)   \\
DU Boo &A   & 8.59  &1.0559 & 0.199 & 0.206  & 0.5343 &	0.2915 & 7850	&7610 & 50 & 81.0 &5.97 &  2.08&  0.49& 	33.42 &    6.32 	&3.19  &1.74  &0.1093 &  51.92  &  (29),(30)    \\
TU Mus  &A  & 8.25  &1.3870 & 0.652 & 0.678& 0.4500 &	0.3700 & 38700&33200& 5  & 77.8 &17.70 & 16.70 &10.40 &104712& 35481  &7.20  &5.70  &0.0444 &  53.91  & (31)          \\
V606 Cen &A & 9.35  &1.4951 & 0.541 & 0.527  & 0.4358 &	0.3281& 29200&21770& 4  & 87.3 &15.58 & 14.70 &7.96 	&30400 &	5440 	&6.83  &5.19  &0.0424 &  53.77  & (32)          \\

V382 cyg &A & 8.33  &1.8855 & 0.677 & 0.742  & 0.4139 &	0.3604 & 36000&34415& 9  & 84.5 &23.45 & 27.90 &20.80 &141905& 89949  &9.70  &8.50  &0.0385 &  54.38  & (33)          \\
LY Aur   &W & 6.68  &4.0020 & 0.528 & 0.528 & 0.4480 &	0.3510& 31000&31150& -  & 87.9 &36.10 & 25.50 &14.00 &213796& 134896 &16.10 &12.60 &0.0461 &  54.31  & (34),(35)     \\
\hline
\end{tabular}
}
\end{center}
(1) \cite{2007AJ....133.1977P}; (2) \cite{2007AJ....133..169D}; (3) \cite{2011MNRAS.412.1787D}; (4) \cite{2008AJ....136..586R}; (5) \cite{2012NewA...17..603E}; (6) \cite{2002AJ....124.1738R}; (7) \cite{2000A&AS..147..243W}; (8) \cite{2005PASJ...57..983Y}; (9) \cite{1993A&A...278..463G}; (10) \cite{2009AJ....137.3655P}; (11) \cite{2014AJ....148..126C}; (12) \cite{2015AJ....149..168P}; (13) \cite{2004AcA....54..299Z}; (14) \cite{2004AJ....127.1712P}; (15) \cite{2005PASJ...57..983Y}; (16) \cite{2021SerAJ.203...29P}; (17) \cite{2018PASP..130c4201L}; (18) \cite{2012NewA...17..143O}; (19) \cite{2005AJ....130..767R}; (20) \cite{2006AcA....56..127G}; (21) \cite{2006AJ....132..769P}; (22) \cite{2001AJ....122.1974R}; (23) \cite{2002AJ....124.1738R}; (24) \cite{2011NewA...16...12C}; (25) \cite{2004A&A...417..725Y}; (26) \cite{2018AJ....156..199S}; (27) \cite{2016AJ....151...67Z}; (28) \cite{2003AJ....125.3258R}; (29) \cite{2013AJ....145...80D}; (30) \cite{2011AN....332..607P}; (31) \cite{2003AJ....126.2988T}; (32) \cite{2022ApJ...924...30L}; (33) \cite{2013AJ....145....9Y}; (34) \cite{2014NewA...26..112Z}; (35) \cite{1994Obs...114..107S}
\end{sidewaystable}

\renewcommand\arraystretch{1.3}
\begin{sidewaystable}
\tiny
\begin{center}
\caption{The long period (P$>$0.5 days) contact binaries without radial velocity observations}
{
\begin{tabular}{p{3cm}p{0.5cm}p{0.5cm}p{0.5cm}p{0.4cm}p{0.4cm}p{0.4cm}p{0.4cm}p{0.4cm}p{0.3cm}p{0.5cm}p{0.5cm}
p{0.4cm}p{0.4cm}p{0.4cm}p{0.4cm}p{0.4cm}p{0.4cm}p{0.8cm}p{0.8cm}p{0.6cm}}
\hline
Star &Type& V$_{max}$&  P &  q$_{ph}$ & r$_{1}$& r$_{2}$& T$_{1}$& T$_{2}$&  f &  i &  a
&  M$_{1}$	&  M$_{2}$&L$_{1}$	 & L$_{2}$	&  R$_{1}$	& R$_{2}$  & J$_{s}$/J$_{o}$
& log J$_{o}$  &Ref.\\
&&(mag)&(days) &&& &(K)&(K)&(\%) &($^{\circ}$) &(R$_\odot$)&(M$_\odot$)
& (M$_\odot$) & (L$_\odot$)& (L$_\odot$)
& (R$_\odot$)&(R$_\odot$)& & (cgs)& \\
\hline
CSS J161753.6+205014&A & 14.60 & 0.5098 &0.080  &0.5980 &0.1930 &7006	&6864&  4   & 78.0  & 6.14 	&1.99&  0.16& 	9.28 	&  0.89 	&  2.07 &0.67& 0.2921 & 51.39 & (1)   \\
BU Vel         &A      & 10.48 & 0.5163 &0.251 &0.5065 &0.2999&7500	&7448&  61  & 84.9  & 10.55 &1.72&  0.43& 	8.90 	&  3.03 	&  1.77 &1.05& 0.0835 & 51.76 & (2)   \\
VV Cet        &A       & 10.54 & 0.5224 &0.249 &0.5100 &0.2600 &8100  &5900&  16  & 80.2  & 3.53 	&1.72&  0.43& 	12.48 &  0.91 	&  1.80 &0.92& 0.0833 & 51.76 & (3)   \\
NSVS 13602901   &A     & 12.03 & 0.5238 &0.171  &0.5540 &0.2600 &6250	&6222&  44  & 83.6  & 9.29 	&1.84&  0.31& 	5.24 	&  1.13 	&  1.96 &0.92& 0.1309 & 51.65 & (4)   \\
CSS J145437.2+060239 &A& 15.28 & 0.5437 &0.110&0.5860 &0.2270 &6930	&6336&  43  & 79.3  & 7.26 	&1.95&  0.21& 	9.34 	&  0.98 	&  2.13 &0.82& 0.2113 & 51.52 & (1)   \\
KIC 8145477    &A      & 14.66 & 0.5658 &0.099 &0.6030 &0.2321 &6538	&6284&  76  & 84.4  & 10.38 &1.98&  0.20& 	8.30 	&  1.05 	&  2.25 &0.87& 0.2462 & 51.49 & (5)   \\
AP Aur        &A       & 11.13 & 0.5694 &0.246 &0.5238 &0.2843&9016	&8703&  64  & 75.9  & 10.53 &1.75&  0.43& 	22.87 &  5.85 	&  1.97 &1.07& 0.0894 & 51.78 & (6)   \\
CSS J021552.4+324419&A& 14.68 & 0.5738 &0.110 &0.5800 &0.2210 &6863	&6315&  25  & 79.3  & 9.42 	&1.97&  0.22& 	9.52 	&  0.99 	&  2.19 &0.83& 0.2069 & 51.53 & (1)   \\
CSS J210300.1+050345&A& 13.90 & 0.5922 &0.100&0.5900 &0.2170&6787	&6188&  34  & 82.8  & 9.17 	&2.00&  0.20& 	9.87 	&  0.92 	&  2.28 &0.84& 0.2329 & 51.51 & (1)   \\
CSS J211420.2-142710&A& 15.32 & 0.5981 &0.110 &0.5780 &0.2180 &6876	&6171&  18  & 74.3  & 9.27 	&1.99&  0.22& 	10.12 &  0.93 	&  2.25 &0.85& 0.2054 & 51.54 & (1)   \\
CSS J234145.7+233158&A& 13.91 & 0.5986 &0.090&0.6100 &0.2260 &6853	&6852&  78  & 77.7  & 10.54 &2.02&  0.18& 	11.14 &  1.53 	&  2.37 &0.88& 0.2737 & 51.47 & (1)   \\
KIC 7601767       &A   & 14.46 & 0.5992 &0.161&0.5507 &0.2464 &6567  &6388&  23  & 77.4  & 3.89  &1.90&  0.31&  7.67   & 1.37    &  2.14 &0.96& 0.1354 & 51.56 & (5)   \\
KIC 7698650     &A     & 14.98 & 0.5992 &0.116 &0.5923 &0.2439&6307	&6261&  74  & 81.8  & 10.72 &1.98&  0.23& 	7.54 	&  1.24 	&  2.31 &0.95& 0.2063 & 51.67 & (5)   \\
NX Cam       &A        & 10.70 & 0.6058 &0.150&0.5639 &0.2548&6657	&6015&  55  & 82.4  & 9.74 	&1.92&  0.29& 	8.62 	&  1.17 	&  2.21 &1.00& 0.1508 & 51.65 & (7)   \\
KIC 8554005 &W         & 12.79 & 0.6083 &0.356 &0.5100 &0.3345&7298	&7302&  59  & 83.3  & 10.23 &1.63&  0.58& 	10.26 &  4.42 	&  2.01 &1.32& 0.0686 & 51.88 & (5)   \\
EK Aqr & A & 11.01 &0.6128 &0.192 & 0.5399 &0.2632 &7900 &6808 & 33 &76.4 &3.96 &1.86&0.36&15.95&2.09&2.14 &1.04 &0.1135&52.43&(8)\\
DN Aur        &W       & 13.00 & 0.6169 &0.205 &0.5414 &0.2765  &6830	&6750&  54  & 76.9  & 11.32 &1.84&  0.38& 	9.05 	&  2.25 	&  2.15 &1.10& 0.1089 & 51.75 & (9)   \\
KIC 5290305       &A   & 14.05 & 0.6210 &0.210 &0.5380 &0.2755&6542	&6186&  41  & 81.6  & 10.66 &1.84&  0.39& 	7.60 	&  1.59 	&  2.15 &1.10& 0.1057 & 51.76 & (5)   \\
KIC 10229723    &A     & 12.04 & 0.6287 &0.141 &0.5681 &0.2457&6477	&6262&  44  & 81.5  & 9.76 	&1.95&  0.28& 	8.29 	&  1.35 	&  2.29 &0.99& 0.1606 & 51.64 & (5)   \\
CSS J234324.8+211100&A& 14.50 & 0.6311 &0.110  &0.5940 &0.2370  &6640	&6603&  68  & 80.6  & 10.32 &2.01&  0.22& 	10.07 &  1.57 	&  2.40 &0.96& 0.2174 & 51.56 & (1)   \\
KIC 11144556        &A & 13.43 & 0.6430 &0.160&0.5824 &0.2847  &6803	&6702&  100 & 76.7  & 11.47 &1.93&  0.31& 	10.97 &  2.47 	&  2.39 &1.17& 0.1532 & 51.69 & (5)   \\
KIC 10007533     &A    & 13.74 & 0.6481 &0.082 &0.6061 &0.2080  &6977	&6379&  44  & 83.4  & 10.28 &2.07&  0.17& 	13.30 &  1.09 	&  2.50 &0.86& 0.2923 & 51.46 & (5)   \\
KIC 3127873      &A    & 15.49 & 0.6715 &0.109&0.6042 &0.2495 &6408	&6164&  98  & 87.5  & 10.97 &2.04&  0.22& 	9.92 	&  1.45 	&  2.56 &1.06& 0.2274 & 51.58 & (5)   \\
IK Per           &A    & 11.15 & 0.6760 &0.191 &0.5482 &0.2724&9070	&7470&  52  & 77.8  & 10.82 &1.91&  0.36& 	33.11 &  3.76 	&  2.34 &1.16& 0.1177 & 51.76 & (10)   \\
KIC 11618883   &W      & 12.44 & 0.6849 &0.225 &0.5523 &0.3045 &4347	&4403&  80  & 87.6  & 12.52 &1.86&  0.42& 	1.81 	&  0.58 	&  2.38 &1.31& 0.1064 & 51.81 & (5)   \\
KIC 9453192     &A     & 13.89 & 0.7188 &0.140 &0.5754 &0.2524 &6622	&6161&  61  & 85.1  & 9.94 	&2.03&  0.28& 	11.39 &  1.64 	&  2.57 &1.13& 0.1666 & 51.68 & (5)   \\
KIC 12352712    &A     & 16.41 & 0.7221 &0.095 &0.6088 &0.2338&6667	&6469&  87  & 89.4  & 10.75 &2.12&  0.20& 	13.19 &  1.72 	&  2.73 &1.05& 0.2594 & 51.55 & (5)   \\
KIC 9350889       &W   & 13.44 & 0.7259 &0.148  &0.5827 &0.2720&6996	&7035&  90  & 87.4  & 11.16 &2.02&  0.30& 	14.77 &  3.29 	&  2.62 &1.22& 0.1628 & 51.71 & (5)   \\
CSS J030702.2+261521&A& 13.98 & 0.7284 &0.090&0.5990 &0.2130 &6680	&6362&  42  & 81.2  & 10.71 &2.13&  0.19& 	13.04 &  1.36 	&  2.70 &0.96& 0.2637 & 51.54 & (1)   \\
KIC 8539720      &A    & 12.93 & 0.7445 &0.148 &0.5828 &0.2717 &6658	&6398&  90  & 82.5  & 11.24 &2.04&  0.30& 	12.60 &  2.34 	&  2.68 &1.25& 0.1632 & 51.72 & (5)   \\
CSS J051156.6+011756&A& 14.82 & 0.7527 &0.150 &0.5600 &0.2450 &6414	&5936&  32  & 83.1  & 10.79 &2.04&  0.31& 	10.20 &  1.43 	&  2.59 &1.13& 0.1484 & 51.72 & (1)   \\
CSS J022044.4+280006&A& 14.42 & 0.7594 &0.150 &0.5630 &0.2490 &6760	&6382&  41  & 84.7  & 10.81 &2.05&  0.31& 	12.89 &  2.00 	&  2.62 &1.16& 0.1501 & 51.73 & (1)   \\
KIC 8265951     &A     & 12.34 & 0.7800 &0.155&0.5601 &0.2509&6943	&6648&  53  & 79.7  & 8.36 	&2.06&  0.32& 	14.80 &  2.50 	&  2.67 &1.19& 0.1449 & 51.75 & (5)   \\
KIC 5439790        &A  & 13.19 & 0.7961 &0.197 &0.5415 &0.2691 &7022	&6804&  39  & 83.0  & 8.28 	&2.00&  0.39& 	14.95 &  3.26 	&  2.62 &1.30& 0.1121 & 51.83 & (5)   \\
V343 Ori           &W  & 10.76 & 0.8091 &0.253 &0.5496 &0.3313&7150	&7312&  87  & 79.7  & 9.65 	&1.92&  0.49& 	16.99 &  6.75 	&  2.69 &1.62& 0.0980 & 51.91 & (11)  \\
V2787 Ori    &A        & 12.70 & 0.8110 &0.120 &0.5717 &0.2225&6993	&5418&  18  & 84.7  & 10.24 &2.15&  0.26& 	16.88 &  0.92 	&  2.81 &1.09& 0.1863 & 51.68 & (12)  \\
KIC 3104113     &W     & 11.50 & 0.8468 &0.171 &0.5777 &0.2897&6535  &6640&  98  & 79.5  & 5.08  &2.09&  0.36&  14.08  & 3.77    & 2.94  &1.45& 0.1430 & 51.81 & (5)   \\
KN Per &A&11.52&0.8665&0.236&0.5712&0.3045&7200&7156&68&85.97&5.18&1.99	&0.49&13.54&	2.43&2.85&	1.28&0.9710&51.93&(13)\\
NS Cam       &A        & 12.78 & 0.9074 &0.213  &0.5242 &0.2619&6250	&5689&  17  & 85.0  & 9.57 	&2.08&  0.44& 	10.84 &  1.86 	&  2.82 &1.41& 0.0989 & 51.91 & (14)  \\
KIC 11097678   &A      & 13.25 & 0.9997 &0.095  &0.6103 &0.2348&6493	&6427&  91  & 83.4  & 10.99 &2.41&  0.23& 	20.05 &  2.85 	&  3.55 &1.37& 0.2625 & 51.69 &  (5)  \\

\hline
\end{tabular}
}
\end{center}
                  (1) \cite{2022MNRAS.tmp..527C}; (2) \cite{1979MNRAS.189..907T}; (3) \cite{2000PASP..112..123R}; (4) \cite{2021RAA....21..235W}; (5) \cite{2021arXiv211214823D}; (6) \cite{2001AJ....121.1091L}; (7) \cite{2021NewA...8401512M}; (8) \cite{2022NewA...9501800G}; (9) \cite{1996Ap&SS.246..291G}; (10) \cite{2005AJ....129.2806Z}; (11) \cite{2017AIPC.1815h0025Y}; (12) \cite{2019PASP..131h4203T}; (13) this paper; (14) \cite{2020JAVSO..48..150S}
\end{sidewaystable}

A statistic of 70 long period (P $>$ 0.5 days) contact binaries containing our target is included in this paper. All collected systems were divided into two parts: the contact binaries with radial velocity observations and the contact binaries without radial velocity observations. For systems with radial velocity observations, the physical parameters are shown in Table 7, containing the type, the brightest V-band magnitude V$_{max}$, period P, photometric mass ratio q$_{ph}$, spectroscopic mass ratio q$_{sp}$, relative radii r$_{1}$ and r$_{2}$, surface temperatures T$_{1}$ and T$_{2}$ of each component, contact degree f, orbital inclination in degree i, the semimajor axis a, the absolute masses M$_{1}$	and M$_{2}$, the absolute luminosity L$_{1}$ and L$_{2}$, the absolute radii R$_{1}$ and R$_{2}$, the spin angular to the orbital angular momentum ratio J$_{s}$/J$_{o}$ and orbital angular momentum log J$_{o}$. For contact binaries without radial velocity observations, the physical parameters shown in Table 8, are the same as Table 7. Among all collected contact binaries, 24 contact binary systems are similar to our target. They are DLMROBs with long period. Because the contact binaries displayed in Table 8 did not have the radial velocity observations, we can not calculate their absolute parameters directly. To estimate their absolute parameters, we obtained the relationship between the period (P) and the semimajor axis (a) by using 26 contact binaries with radial velocity observations. Because the period range of contact binaries without radial velocity observations collected in this paper is from 0.5 days to 1 days, we used contact binaries with period near and less than 1 days to determine the relation of P-a. A linear equation was used to fit it, which is shown as follows:

\begin{equation}
a = 1.02(\pm0.32) + 4.79(\pm0.45)P.
\end{equation}
The relation of P-a is shown in Figure 7. The RMS of this relation is 0.68. To verify the accuracy of Equation (5), we recalculated the semimajor axis of the binaries with radial velocity observations by the Equation (5). Then, we obtained the differences between the calculated semimajor axis and the original semimajor axis. The ratios of the differences and the the semimajor axis $\Delta$$\alpha$ were determined. There are 74\% systems with $\Delta$$\alpha$ less than 10\%. Additionally, we calculated the semimajor axis of the binaries without radial velocity observations using the above equation. R$_{1}$ and R$_{2}$ were obtained by the relative radii and the semimajor axis. Then, according to the third law of Kepler: M$_1$ (M$_\odot$) + M$_2$ = 0.0134a$^3$ /P$^2$ and M$_{2}$ = M$_{1}$ $\times$ q, the masses of the two components were determined. The units of M, a and P are M$_\odot$, R$_\odot$ and d. Finally, we obtained L$_{1}$ and L$_{2}$ by L = 4$\pi$$\sigma$T$^{4}$R$^{2}$. We show the absolute parameters of the binaries without radial velocity observations in Table 8, containing the semimajor axis a, the absolute masses M$_{1}$ and M$_{2}$, the absolute luminosity L$_{1}$ and L$_{2}$, the absolute radii R$_{1}$ and R$_{2}$. The uncertainties of all absolute parameters of KN Per were obtained by using the error propagation, which are displayed in Table 9.

\renewcommand\arraystretch{1.3}
\begin{table}
\small
\caption{The uncertainties of the absolute parameters of KN Per}
\begin{tabular}{p{2cm}p{2cm}p{2cm}p{2cm}p{2cm}p{2cm}p{2cm}}
\hline
a        & M$_{1}$           & M$_{2}$   & L$_{1}$ & L$_{2}$ & R$_{1}$ & R$_{2}$ \\
(R$_\odot$)&(M$_\odot$)
& (M$_\odot$) & (L$_\odot$)& (L$_\odot$)
& (R$_\odot$)&(R$_\odot$) \\
\hline
0.71 &0.83&0.20&0.41&0.22&2.40&0.58\\
\hline
\end{tabular}
\end{table}

\begin{figure}[h]
\centering
\includegraphics[width=14cm]{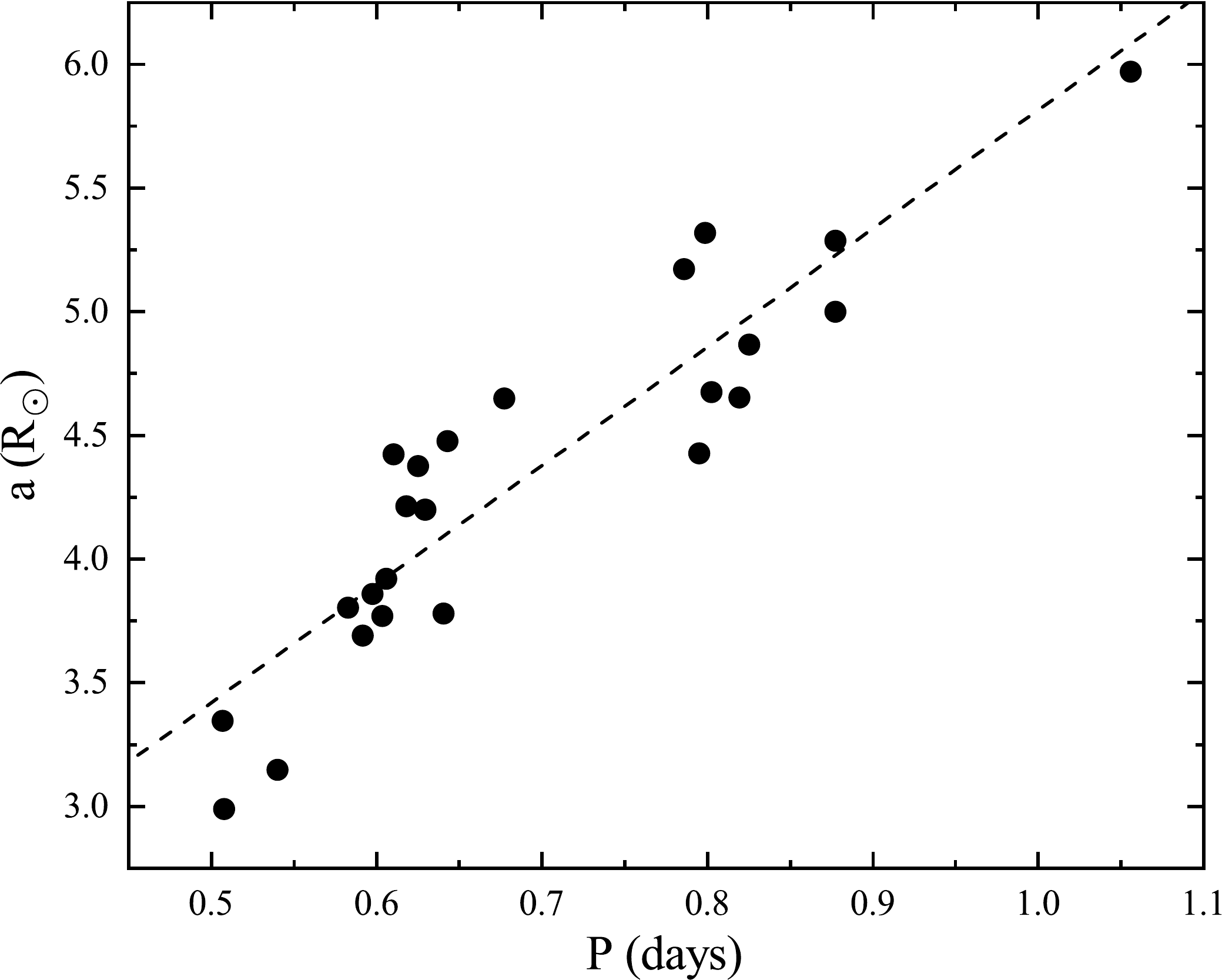}
\caption{The relationship between the period (P) and the semimajor axis (a). }
\end{figure}

The relations of q-R$_{2}$/R$_{1}$ and q-f are displayed in Figure 8 and two equations were used to fit them. The range of the ratio of secondary radius to primary radius is 0.3 - 1.0. The equation showing the relation of q-R$_{2}$/R$_{1}$ is displayed as follows:

\begin{equation}
R_{2}/R_{1} = 0.27(\pm0.01) + 1.26(\pm0.09)q -0.61(\pm0.11)q^{2}.\\
\end{equation}
From the left panel of Figure 8, barely binary systems are deviated from the general tendency of this equation. From the right panel of Figure 8, the locations of the contact binaries with small mass ratio (q $\leq$ 0.3) are distributed and the contact degree of them are from 0 $\%$ to 100 $\%$ and the contact degree of the systems with mass ratio more than 0.3 are smaller than 70 $\%$ and decrease with the increasing mass ratio. A nonlinear equation was applied to fit it,

\begin{equation}
f = 25.56(\pm28.82) + 41.94(\pm21.31)e^{-2.96(\pm1.76)q}.
\end{equation}

\cite{2022MNRAS.512.1244C} found that the smallest the q value, the larger the contact degree distribution range. Our result is similar to the study of them.

\begin{figure}\centering
\includegraphics[width=0.45\textwidth]{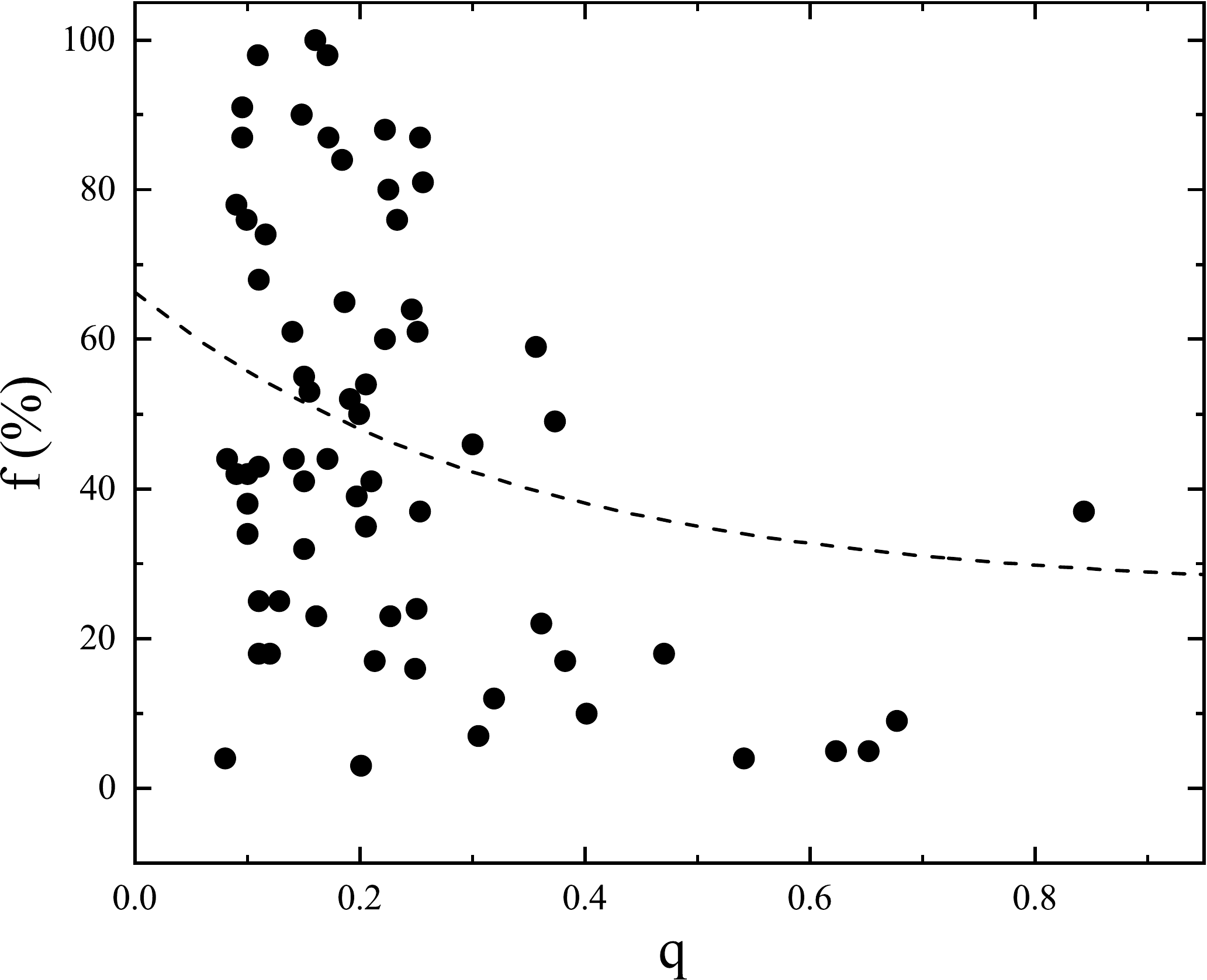}
\includegraphics[width=0.45\textwidth]{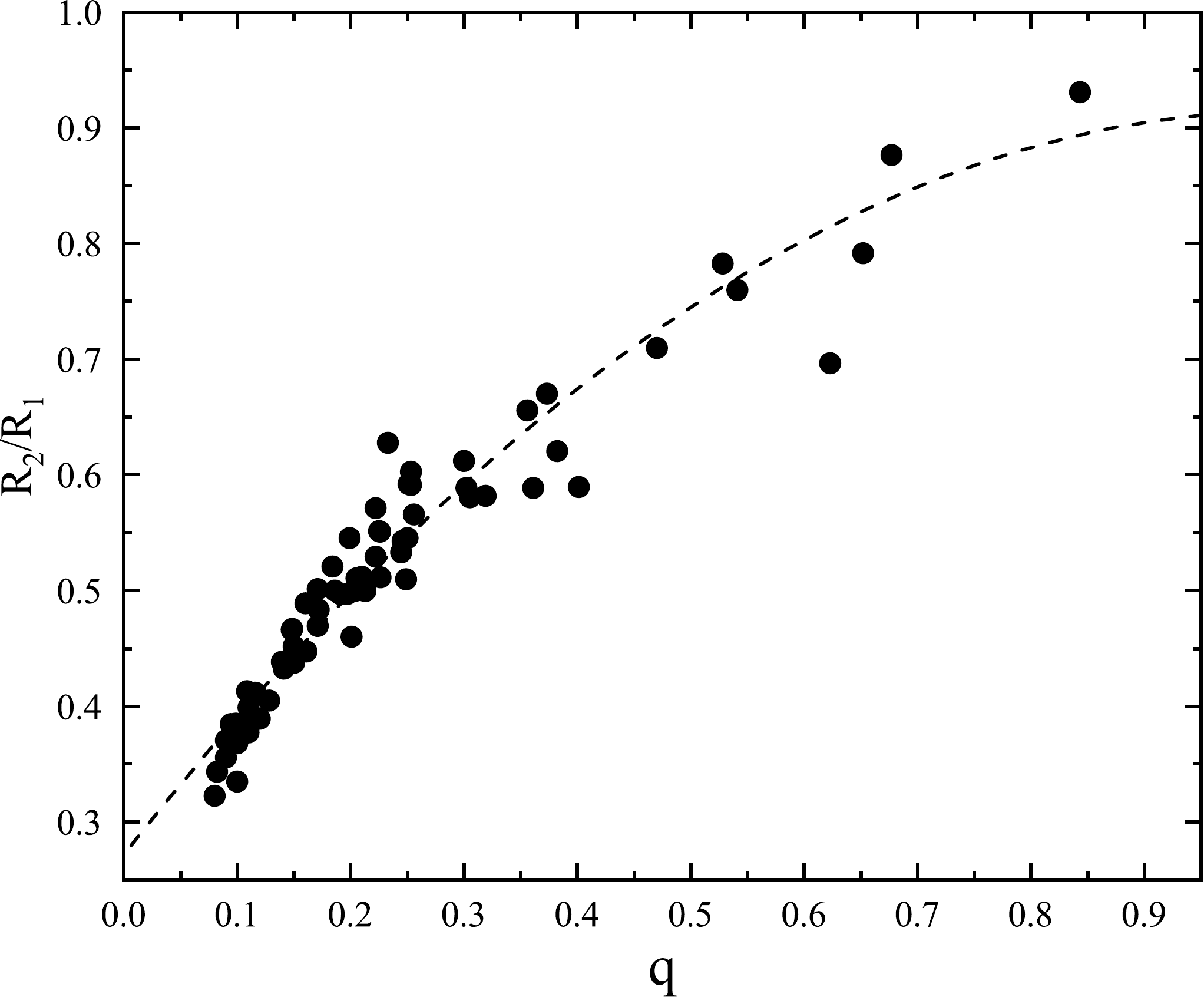}
\caption{The relations of q-R$_{2}$/R$_{1}$ (left) and q-f (right).}
\end{figure}

With k$_{1,2}^2$ fixed on 0.06 \citep{2006MNRAS.369.2001L}, the ratios of the spin angular momentum to the orbital angular momentum (J$_{s}$/J$_{o}$) were calculated by the following equation from \cite{2015AJ....150...69Y}:

\begin{equation}
\frac{J_{s}}{J_{o}} = \frac{1+q}{q}[(k_1r_1)^2+(k_2r_2)^2q], \\
\end{equation}
where r$_{1,2}$ refer to the relative radii and k$_{1,2}^2$ are the dimensionless gyration radii. We display the relation of q-J$_{s}$/J$_{o}$ of all collected systems in Figure 9 and the dashed line is the boundary to distinguish stable boundary and unstable boundary. It was obvious that the value of J$_{s}$/J$_{o}$ falls with the increasing mass ratio. We used an exponential equation to fit it:

\begin{equation}
\frac{J_{s}}{J_{o}} = 0.050(\pm0.002) + 0.602(\pm0.019) \times e^{-11.581(\pm0.353) \times q}.\\
\end{equation}
 The predicted minimum is the mass ratio corresponding to Js/Jo=1/3. A predicted minimum mass ratio was calculated as 0.0652. The minimum mass ratio was determined as q$_{min}$ $\sim$ 0.076 when considering the rotation of the secondary component and the dimensionless gyration radii k$_{1}$$^{2}$ = k$_{2}$$^{2}$ = 0.06 \citep{2006MNRAS.369.2001L}. The value was derived as 0.070 $\thicksim$ 0.074 when considering the rotation effect \citep{2007MNRAS.377.1635A, 2009MNRAS.394..501A}. A predicted minimum mass ratio was calculated as 0.0652. Our result is consistent with many previous studies \citep{2006MNRAS.369.2001L, 2021ApJ...922..122L}.

\begin{figure}[h]
\centering
\includegraphics[width=14cm]{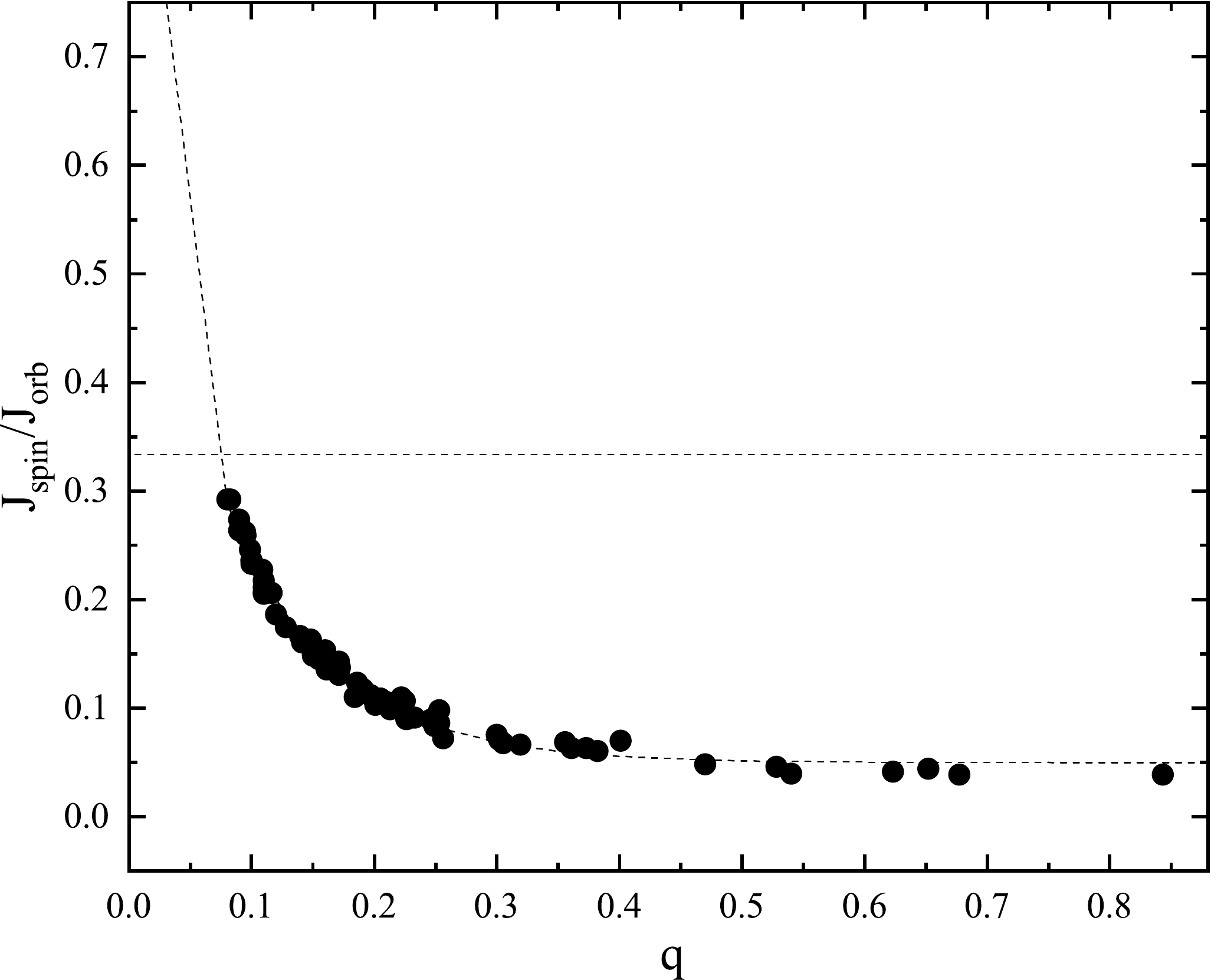}
\caption{The q-J$_{s}$/J$_{o}$ diagram, where the dashed line refers to the boundary distinguishing stable boundary and unstable boundary.}
\end{figure}

We divided all collected contact binaries into two types: A-type and W-type. With the parameters displayed in Table 7 and Table 8, the relations of mass-radius (M-R) and mass-luminosity (M-L) are shown in Figure 10. According to \cite{2002MNRAS.329..897H}, the zero-age main-sequence (ZAMS) and the terminal-age main-sequence (TAMS) are determined from the binaries evolution code, which are displayed as the solid and dashed lines in Figure 10. We use the circles to represent the two components of A-type contact binaries, the triangles to refer to the two components of W-type contact binaries, and the two red circles to represent the two components of KN Per. The open symbols refer to the more massive stars of A-type contact binaries and the solid symbols refer to the less massive stars of A-type binaries. For W-type contact binaries, The open symbols and the solid symbols also refer to the more massive stars and the less massive stars, respectively. The locations of the two components in KN Per are in accord with those of other stars. For both A-type contact binaries and W-type contact binaries, most of the primary components are situated between the lines of "ZAMS" and "TAMS", signifying that they are main sequence stars. While most of the secondary components of both A-type contact binaries and W-type contact binaries are situated above the line of "TAMS", implying that they have evolved from the main sequence stage. Contrast to the secondary components, the primary components are located near the line of "ZAMS". It indicates that the evolutionary stages of the primary components are earlier than those of the secondary components, the reason of which may be the energy and mass transfer from the primary components to the secondary components. The result obtained by us supports the result obtained by previous researchers \citep{2005ApJ...629.1055Y, 2017RAA....17...87Q, 2018ApJS..235....5Q}.

\begin{figure}\centering
\includegraphics[width=0.45\textwidth]{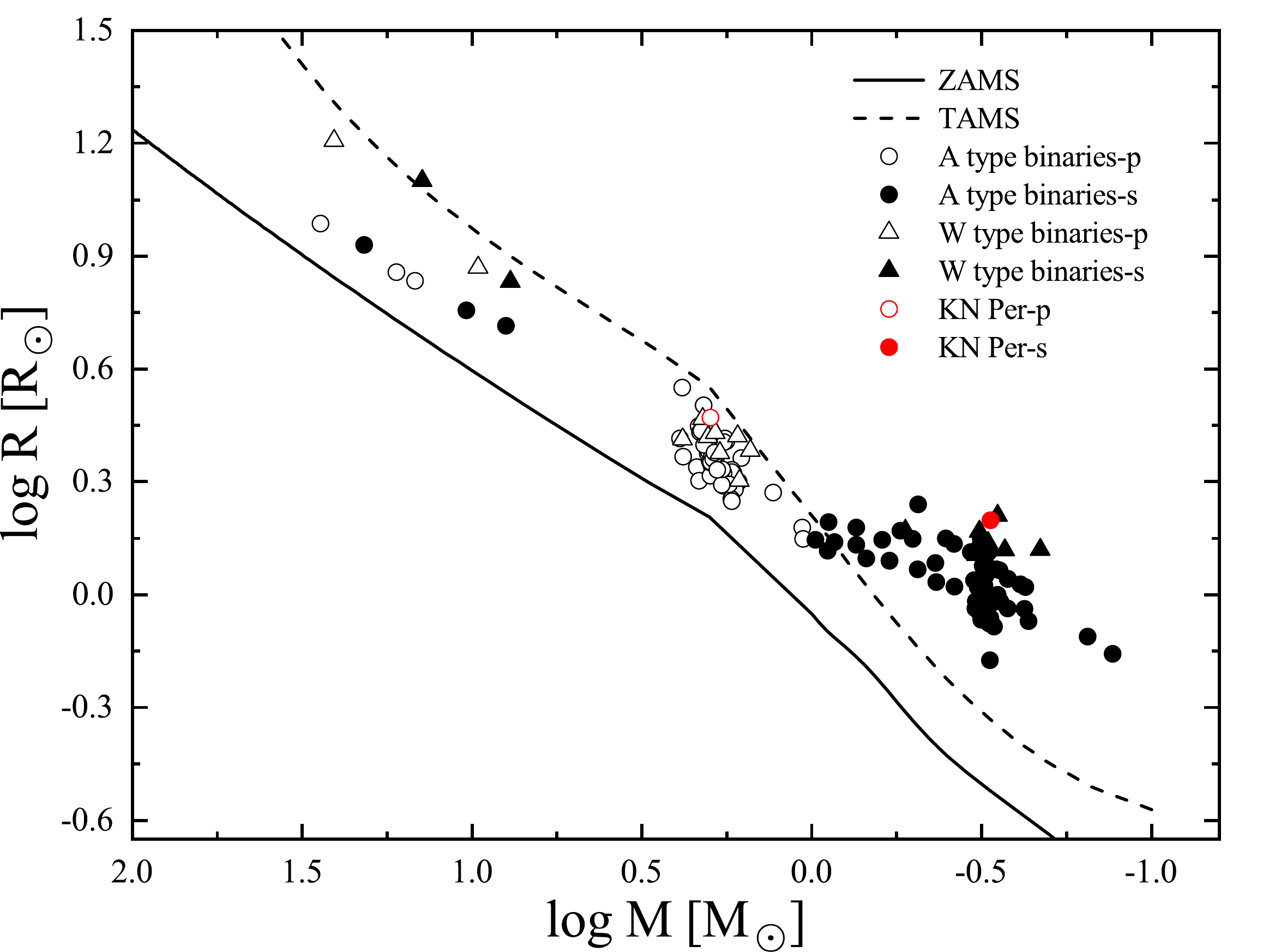}
\includegraphics[width=0.45\textwidth]{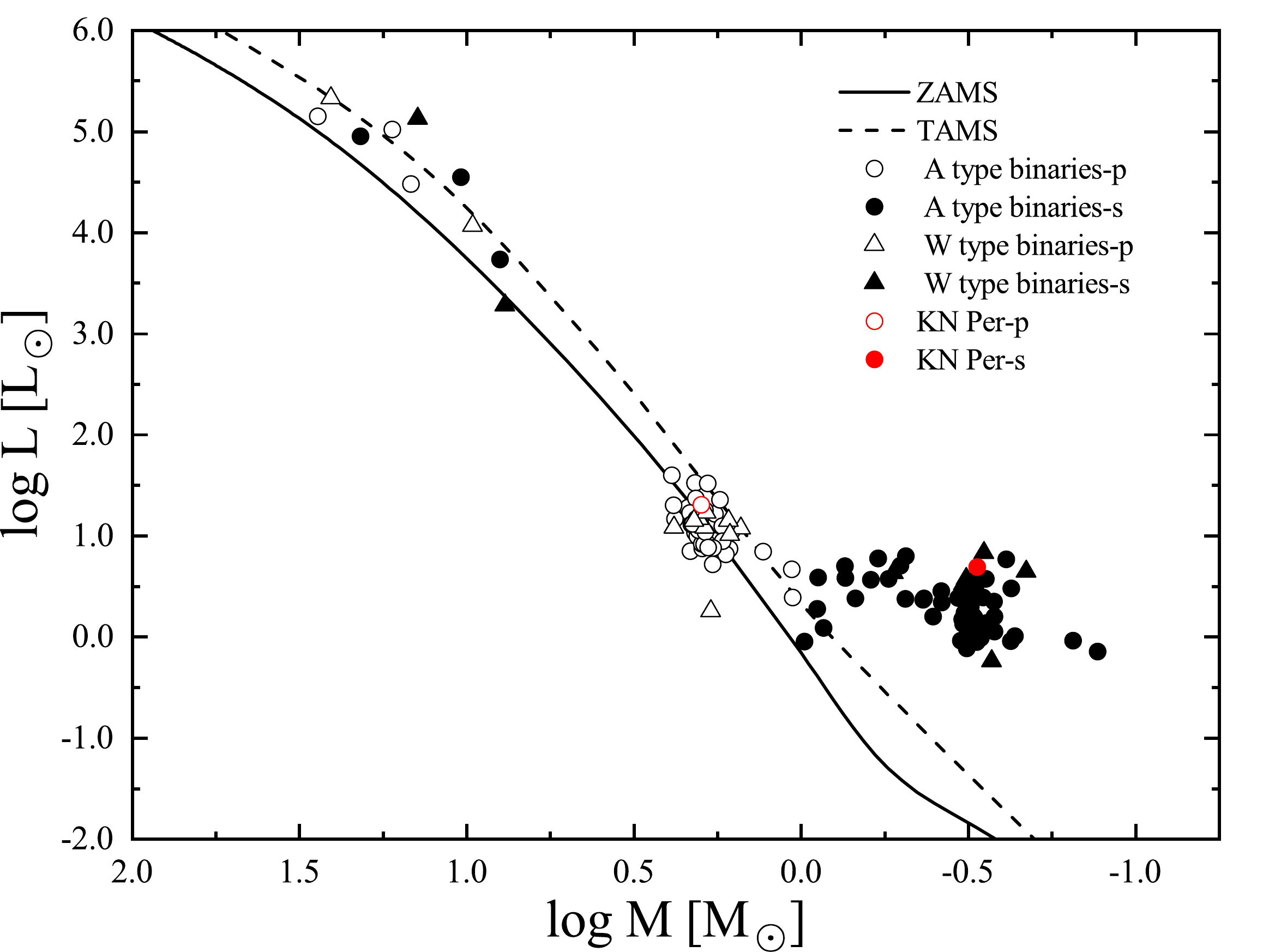}
\caption{The mass-radius diagram (left) and the mass-luminosity diagram (right). The hollow and solid circles represent the A-type contact binaries, the open and solid triangles represent the W-type contact binaries, and the two red circles are two components of KN Per.}
\end{figure}

With the total mass of the binary M$_{T}$ (solar unit), P (days) and q, we calculated the orbital angular momentum J$_{o}$ of the collected contact binaries, based on the following equation from \cite{2013AJ....146..157C}:

\begin{equation}
J_{o} = 1.24 \times 10^{52} \times M^{3/5}_T \times P^{1/3} \times q \times (1+q)^{-2}.
\end{equation}
The relation of log M$_{T}$ - log J$_{o}$ is shown in Figure 11, containing the detached binary systems and the boundary line. The boundary segregates detached and contact binaries \citep{2006MNRAS.373.1483E}. Nearly all contact binary systems are located below the boundary, while the detached binaries are situated above it. This means that J$_{o}$ of contact binary systems are less than those of detached binary systems with the same mass. This is due to the angular momentum loss while contact binaries were forming and evolving. To some extent, this shows that the formation of DLMROBs is initially from the detached binary systems with short periods and DLMROBs are formed through angular momentum loss via magnetic stellar wind \citep{2007AJ....134.1769Q, 2013ApJS..207...22Q, 2015ApJ...798L..42Q, 2018ApJS..235....5Q, 2019MNRAS.485.4588L}.

\begin{figure}[h]
\centering
\includegraphics[width=14cm]{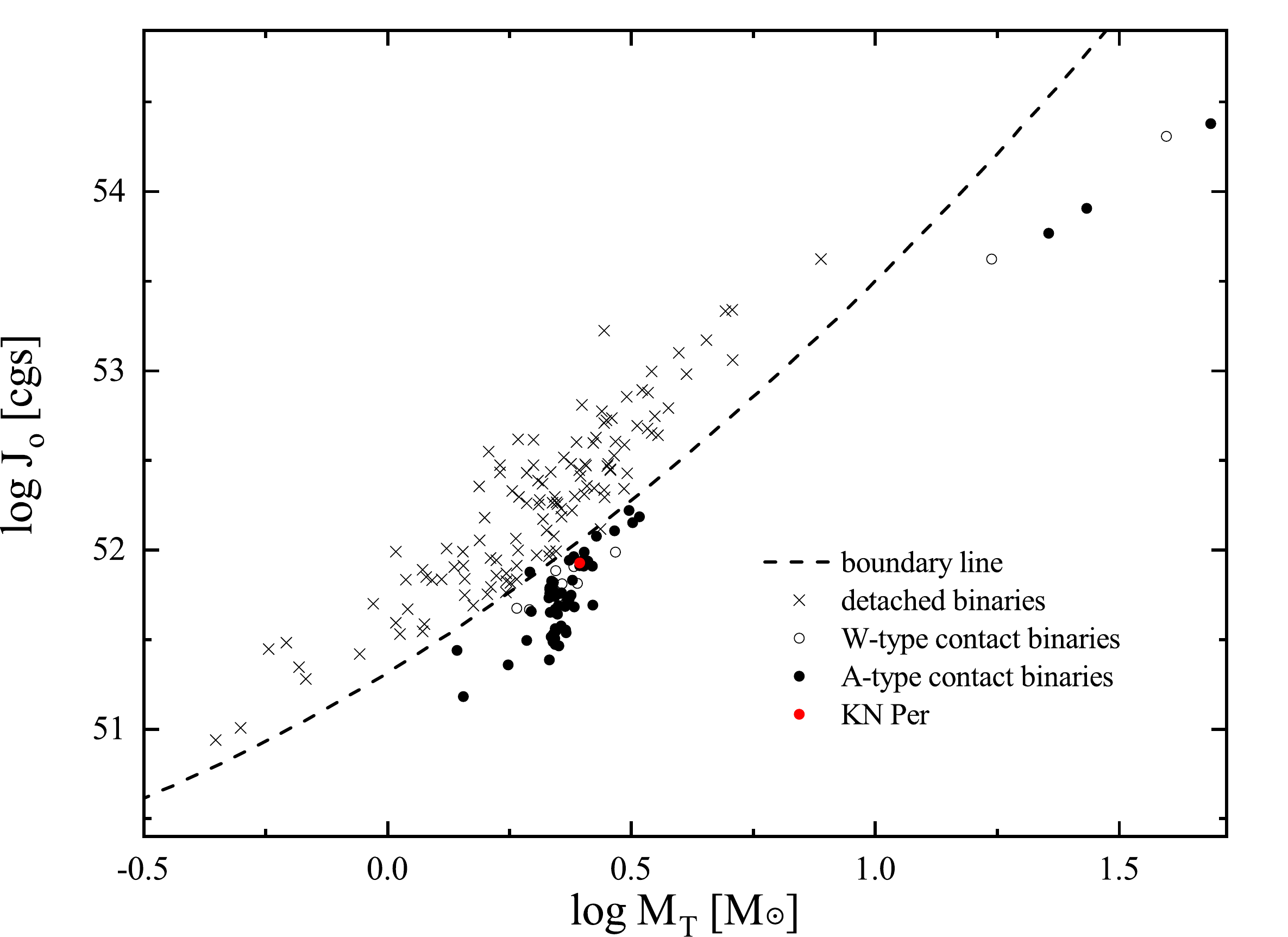}
\caption{The relationship between orbital angular momentum and total mass for binary systtems. The black crosses represent the detached binaries. The hollow circles represent the A-type contact binaries and the solid circles refer to the W-type contact binaries. KN Per is shown as the red circle.}
\end{figure}

To further explore the stability of KN Per, the instability parameters were calculated through the equations provided by \cite{2021MNRAS.501..229W}. Firstly, based on the mass of the primary component and the contact degree, we obtained the instability mass ratio q$_{inst}$ as 0.0428. Then, with the photometric mass ratio, the contact degree and the semimajor axis, the theoretical fractional radius of the primary component was calculated, which is 2.815 R$_\odot$. We determined the dimensionless gyration radius of the primary component (k$_{1}$ = 0.1801) for KN Per (M$_{1}$ = 2.0 M$_\odot$) from \cite{2009A&A...494..209L}. The value of k$_{2}$ was also obtained from this literature. With R$_{1}$, k$_{1}$, k$_{2}$ (k$_{2}^{2}$=0.205) and q, we determined that the instability separation A$_{inst}$ is 3.1042 R$_\odot$. Finally, the instability period was obtained to be 0.4025 days, according to the third law of Kepler. Because the values of q$_{inst}$, A$_{inst}$ and P$_{inst}$ are far less than the three corresponding actual parameters, KN Per is in a stable stage.

We calculated the material transfer rate using the following equation,

\begin{equation}
\frac{dM_{1}}{dt} = \frac{M_{1}M_{2}}{3P(M_{1}-M_{2})} \times\ \frac{dP}{dt}.\\
\end{equation}
The rate is 1.28 $\times$ 10$^{-7}$ M$_\odot$/yr. For KN Per,the rate of period variation and mass transfer is typical for W UMa systems. The rate of period variation and mass transfer of five contact binaries are displayed in Table 10. The value is positive, showing the mass transfer from the secondary component to the primary component. It will lead to a falling mass ratio. In the future, the mass ratio of KN Per may reach the dynamic stability limit and KN Per may merge into a fast rotating star.

\begin{table}
\begin{center}
\small
\caption{The rate of period variation and mass transfer of five contact binaries}
\begin{tabular}{p{3cm}p{4cm}p{4cm}p{3cm}}
\hline
Name        & The rate of period variation           & The rate of the mass transfer   & Ref. \\
&(d yr$^{-1}$)
& (M$_\odot$ yr$^{-1}$) & \\
\hline
V410 Aur& 8.22 $\times$ 10$^{-7}$& 1.62 $\times$ 10$^{-7}$ &(1)\\
XY Boo& 6.25 $\times$ 10$^{-7}$& 1.17 $\times$ 10$^{-7}$&(1)\\
V1191 Cyg& 4.50 $\times$ 10$^{-7}$& 7.45 $\times$ 10$^{-8}$& (2)\\
TY Pup &5.57 $\times$ 10$^{-8}$&8.41 $\times$ 10$^{-9}$&(3)\\
FN Cam& 4.38 $\times$ 10$^{-7}$ &1.47 $\times$ 10$^{-7}$& (4)\\
\hline
\end{tabular}
\end{center}
                  (1) \cite{2005AJ....130.2252Y}; (2) \cite{2011AJ....142..124Z}; (3) \cite{2018AJ....156..199S}; (4) \cite{2018NewA...62...20H}
\end{table}

In conclusion, KN Per is an A-type long period DLMROB and has weak chromosphere activity due to the photometric and spectroscopic analysis. Its orbital period exhibits a continuous increase with a rate of 5.12 $\pm$ (0.30) $\times$ 10$^{-7}$ d/yr. We also found that KN Per is in a stable stage by calculating the instability parameters and shows a mass transfer from the secondary component to the primary component.

\begin{acknowledgements}
This work is supported by the Joint Research Fund in Astronomy (No. U1931103) under cooperative agreement between National Natural Science Foundation of China (NSFC) and Chinese Academy of Sciences (CAS), and by NSFC (No. 1227301811703016), and by the Natural Science Foundation of Shandong Province (No. ZR2014AQ019), and by Young Scholars Program of Shandong University, Weihai (No. 20820171006), and by the Open Research Program of Key Laboratory for the Structure and Evolution of Celestial Objects (No. OP201704), and by the Cultivation Project for LAMOST Scientific Payoff and Research Achievement of CAMS-CAS.

This work is partly supported by the Supercomputing Center of Shandong University, Weihai.

We acknowledge the support of the staff of NEXT.

This work includes data collected by the TESS mission. Funding for the TESS mission is provided by NASA Science Mission directorate. We acknowledge the TESS team for its support of this work.

This publication makes use of data products from the AAVSO Photometric All Sky Survey (APASS). Funded by the Robert Martin Ayers Sciences Fund and the National Science Foundation.

We thank Las Cumbres Observatory and its staff for their continued support of ASAS-SN. ASAS-SN is funded in part by the Gordon and Betty Moore Foundation through grants GBMF5490 and GBMF10501 to the Ohio State University, and also funded in part by the Alfred P. Sloan Foundation grant G-2021-14192.

The spectral data were provided by Guoshoujing Telescope (the Large Sky Area Multi-Object Fiber Spectroscopic Telescope LAMOST) is a National Major Scientific Project built by the Chinese Academy of Sciences. Funding for the project has been provided by the National Development and Reform Commission. LAMOST is operated and managed by the National Astronomical Observatories, Chinese Academy of Sciences.

This paper makes use of data from the DR1 of the WASP data (\citealt{2010A&A...520L..10B}) as provided by the WASP consortium,
and the computing and storage facilities at the CERIT Scientific Cloud, reg. no. CZ.1.05/3.2.00/08.0144
which is operated by Masaryk University, Czech Republic.

\end{acknowledgements}

\bibliography{bib}
\bibliographystyle{aasjournal}

\end{document}